\documentclass[prb,twocolumn,aps,showpacs,eqsecnum,amsmath,
superscriptaddress]{revtex4}
\usepackage{graphicx,amsmath,amssymb,bm,times}

\begin{document}

\title{
Ground-state phase diagram of the one-dimensional half-filled extended Hubbard model
}

\author{M. Tsuchiizu}
\affiliation{Yukawa Institute for Theoretical Physics, 
         Kyoto University, Kyoto 606-8502, Japan}
\affiliation{Department of Physics, Nagoya University,
         Nagoya 464-8602, Japan}
\author{A. Furusaki}
\affiliation{Yukawa Institute for Theoretical Physics, 
         Kyoto University, Kyoto 606-8502, Japan}
\affiliation{Condensed-Matter Theory Laboratory,
    The Institute of Physical and Chemical Research (RIKEN),
         Saitama 351-0198, Japan}
\date{January 26, 2004}

\begin{abstract} 
We revisit the ground-state phase diagram of the one-dimensional
   half-filled extended Hubbard model with on-site ($U$) and 
   nearest-neighbor ($V$) repulsive interactions.
In the first half of the paper, using the weak-coupling
   renormalization-group approach ($g$-ology) 
   including second-order corrections to the coupling constants,
   we show that bond-charge-density-wave (BCDW)
   phase exists for $U\approx 2V$ in between
   charge-density-wave (CDW) and spin-density-wave (SDW) phases.
We find that the umklapp scattering of parallel-spin electrons
   disfavors the BCDW state and leads to a bicritical point
   where the CDW-BCDW and SDW-BCDW continuous-transition lines merge
   into the CDW-SDW first-order transition line.
In the second half of the paper, we investigate
   the phase diagram of the extended Hubbard model with either
   additional staggered site potential $\Delta$
   or bond alternation $\delta$.
Although the alternating site potential $\Delta$ strongly favors
   the CDW state (that is, a band insulator), the BCDW state is not
   destroyed completely and occupies a finite region
   in the phase diagram.
Our result is a natural generalization
   of the work by Fabrizio, Gogolin, and Nersesyan
   [Phys.\ Rev.\ Lett.\ \textbf{83}, 2014 (1999)], who predicted the
   existence of a spontaneously dimerized insulating state between
   a band insulator and a Mott insulator in the phase diagram of
   the ionic Hubbard model.
The bond alternation $\delta$ destroys the SDW state and changes it 
   into the BCDW state (or Peierls insulating state).
As a result the phase diagram of the model with $\delta$ contains
   only a single critical line
   separating the Peierls insulator phase and the CDW phase.
The addition of $\Delta$ or $\delta$ changes the universality class of 
   the CDW-BCDW transition from the Gaussian transition
   into the Ising transition.
\end{abstract}


\maketitle

\section{Introduction}

It is well known that 
  a one-dimensional (1D) spin system has instability to dimerization
  that changes the system into a nonmagnetic insulating state,
  the so-called spin-Peierls state. \cite{Bray}
Indeed the spin-Peierls state is realized in many systems including
   quasi-one-dimensional organic compounds \cite{Ishiguro,Kagoshima} 
   and the inorganic material\cite{Hase} CuGeO$_3$,
   and its properties have been studied extensively 
   both experimentally and theoretically.
Of particular interest is a situation in which a dimerized state
   appears spontaneously due to strong correlations
   and frustration.\cite{Sachdev}
A well-known example is the frustrated spin-$\frac{1}{2}$ Heisenberg
   chain with nearest-neighbor, $J_1$, and next-nearest-neighbor, $J_2$, 
   antiferromagnetic exchange interactions, where
   a spontaneously dimerized phase is realized
   for $J_2 \ge J_{2c}\simeq 0.24 J_1$.\cite{Majumdar_Ghosh}
Other systems of current interest are quasi-one-dimensional electron
   systems in organic materials, where the spin-Peierls state appears
   due to strong electron correlation at
   half filling
   \cite{Ukrainskii,Kivelson,Mazumdar,Hirsch83,Hara,%
    Sugiura2002,Malek,Sengpta2003} and 
   at quarter filling.\cite{Kuwabara,Sugiura2003}

Recently it was pointed out by Nakamura\cite{Nakamura} and co-workers that
   a spontaneously dimerized state occupies a finite parameter space
   in the ground-state phase diagram of the 1D half-filled 
   Hubbard model with the nearest-neighbor repulsion $V$, i.e.,
   the extended Hubbard model (EHM).
This spin-Peierls state is often called
   bond-charge-density-wave (BCDW) state or bond-ordered-wave state.
The appearance of the BCDW state in the purely electronic model is
   nontrivial and has attracted much attention from theoretical
   point of view.
To appreciate this surprising result, let us consider
   some limiting cases.
In the limit of weak nearest-neighbor repulsion $V$, or 
   in the half-filled Hubbard model with only the on-site Coulomb
   repulsion $U$, the ground state is in the Mott insulating state
   where the spin sector exhibits quasi-long-range order of
   spin-density wave (SDW); we call it the SDW state.
In the opposite limit of strong $V$,
   the ground state of the half-filled EHM has a long-range
   order of the charge-density wave (CDW);
   we call this state the CDW state.
Furthermore, in the atomic limit where the electron hopping $t$
  is ignored,
  the CDW state appears for $U<2V$ whereas the uniform
  state corresponding to the SDW state is stable for $U>2V$
  in one dimension.
Strong-coupling perturbation theory in $t$ has established
  that a first-order phase transition
  between the SDW state and the CDW state occurs at $U\simeq 2V$.
  \cite{Emery,Bari,Hirsch,Dongen}
As for the weak-coupling regime, perturbative renormalization-group
  (RG) approach or $g$-ology led to a similar conclusion that
   the ground state at half filling is either in the SDW state
   or in the CDW state with a continuous phase-transition line
   at $U=2V$.\cite{Emery}
Thus, it had been considered for a long time that
   the ground-state phase diagram of the EHM at half filling has
   only two phases, the SDW and CDW states, and that the order of
   the phase transition at $U\simeq2V$ changes from continuous to
   first order at a tricritical point which was speculated to exist
   in the intermediate coupling regime. 
   \cite{Hirsch,Cannon,Zhang,Voit}
This common view was revised by the Nakamura's discovery that
   the BCDW state exists at $U\simeq 2V$
   in between the SDW and CDW phases
   in the weak-coupling region,\cite{Nakamura}
   which is supported by recent large-scale Monte Carlo
   calculations.\cite{Sengupta,Sandvik}
Related studies of the dimerized state in
   the EHM with additional correlation effects can be found in
  Refs.\ [\onlinecite{Fukutome,Japaridze,Otsuka,Aligia,Nakamura2001}].

A related and still controversial issue of current interest is
   whether or not a spontaneously dimerized phase exists
   in the 1D Hubbard model with alternating site
   potential, the so-called ionic Hubbard
   model.\cite{Fabrizio,Tsuchiizu_JPSJ,Takada,Qin,Brune,YZZhang,Anusooya-Pati,%
   Wilkens,Torio,Refolio,Caprara,Gupta,Pozgajcic,Manmana}
This system was introduced as a simple minimal model for
    the neutral-ionic transitions observed in 
    quasi-one-dimensional organic materials\cite{MJRice,Nagaosa,Girlando}
   and for ferroelectric perovskites.\cite{Egami,Resta}
Obviously the model has two insulating phases.
The ground state is (i) a band insulator with the CDW order when
   the staggered site potential is much larger than the on-site
   repulsion or
   (ii) a Mott insulator with quasi-long-range SDW order
   when the staggered site potential is negligible.
Early exact diagonalization studies\cite{Egami,Resta,Gidopoulos}
   of small systems have found a transition between the two
   phases and also reported dramatic enhancement of the
   electron-lattice interaction
   by strong electron correlation near a boundary between
   the band insulating phase (the BI state) and the Mott
   insulating phase (the SDW state).
Mostly through bosonization analysis of the ionic Hubbard model,
   Fabrizio, Gogolin, and Nersesyan recently argued\cite{Fabrizio}
   that a phase of a spontaneously dimerized insulator (SDI)
   intervenes between the ionic insulating phase (band insulator)
   and the Mott insulating phase.
The SDI state is closely related to the BCDW state mentioned above.
Earlier numerical studies
  \cite{Takada,Qin,Brune,Anusooya-Pati,Wilkens,Gidopoulos}
   have drawn contradictory conclusions as to whether
   the SDI phase exists or not, but more recent numerical studies
   find two phase transitions and the SDI phase
   in between.\cite{Torio,YZZhang,Manmana}
Nevertheless there still remain unresolved issues on
   the critical properties near the quantum phase transitions.

In this paper we give supporting theoretical arguments for the existence
   of the spontaneously dimerized insulating states in
   the 1D half-filled extended Hubbard model with and without
   staggered potentials.
We adopt the standard bosonization approach and perform both perturbative
   RG analysis valid in the weak-coupling regime and semiclassical
   analysis which is expected to give a qualitatively correct picture
   even in the strong-coupling regime.
This paper is organized as follows.
Sections \ref{sec:model} and \ref{sec:phase_diagram} are 
   devoted to the analysis of the standard EHM, i.e., the system 
   without the staggered potential.
Some of the results of this part are already presented 
   in Ref.\ \onlinecite{Tsuchiizu_Furusaki}.
In Sec.\ \ref{sec:model},
   we introduce the model and 
   reformulate the weak-coupling theory, the $g$-ology,
   to include higher-order corrections to coupling constants.
We bosonize low-energy effective Hamiltonian
   and derive the renormalization-group equations.
In Sec.\ \ref{sec:phase_diagram},
   we determine the ground-state phase diagram. 
First, from the perturbative RG analysis we show that 
   the BCDW phase occupies a finite region near the $U=2V$ line
   in the weak-coupling limit.
Next, from the semiclassical analysis
   we argue that the umklapp scattering 
   of parallel-spin electrons
   destabilizes the BCDW phase and gives rise to a bicritical point
   where the CDW-BCDW and SDW-BCDW continuous-transition lines merge
   into the CDW-SDW first-order transition line.
Finally,
   combining the perturbative RG equations with the semiclassical
   analysis, we obtain the global phase diagram of the 1D EHM.
In Sec.\ \ref{sec:staggered}
   we study the 1D EHM with the staggered site potential.
We take the same strategy as in the previous sections and
   perform a semiclassical analysis of the bosonized Hamiltonian.
With the help of the perturbative RG analysis
   we obtain the global phase diagram that indeed has the SDI phase.
We find that the BCDW phase of the EHM is continuously deformed to
   the SDI phase
   upon introducing the alternating site potential.
In Sec.\ \ref{sec:dimer}, we study the 1D EHM with additional bond
   dimerization, but without the staggered potential.
This model exhibits a quantum phase transition between a dimerized
   Peierls insulator and a CDW state.
Section \ref{sec:conclusions} is devoted to conclusions,
   and details of the technical calculations are given in 
   Appendixes.

\section{Extended Hubbard Model}\label{sec:model}

In the first half of this paper 
   (Secs.\ \ref{sec:model} and \ref{sec:phase_diagram}),
   we consider the standard 1D EHM which has on-site, $U$, and
   nearest-neighbor, $V$, interactions.
The Hamiltonian is given by
\begin{eqnarray}
H &=&
 - t \sum_{j,\sigma}
   ( c_{j,\sigma}^\dagger c_{j+1,\sigma}^{} + \mathrm{H.c.} )
\nonumber \\
&&{}
 + U \sum_j n_{j,\uparrow} \, n_{j,\downarrow}
   + V \sum_j n_{j} \, n_{j+1} ,
\label{eq:H1D}
\end{eqnarray}
  where
  $n_{j,\sigma} \equiv c_{j,\sigma}^\dagger c_{j,\sigma}^{}
   -\frac{1}{2}$,
  $n_j \equiv n_{j,\uparrow} + n_{j,\downarrow}$, and
  $c_{j,\sigma}^\dagger$ denotes the creation operator of an
  electron with spin $\sigma$ 
  ($=\,\uparrow$, $\downarrow$) on the $j$th site.
We assume repulsive interactions, i.e.,
   the coupling constants $U$ and $V$ are positive.
Note that the Hamiltonian has global SU(2) spin symmetry.
Following the previous studies on models with correlated-hopping
  interactions,\cite{Japaridze} we consider the CDW, SDW, BCDW, and 
  bond-spin-density-wave (BSDW) phases as potential ordered ground
  states at half filling.
They are characterized by the order parameters
\begin{subequations}
\begin{eqnarray}
\mathcal{O}_\mathrm{CDW} \!\!&\equiv&\!\! (-1)^j \,
   (n_{j,\uparrow}+n_{j,\downarrow})
,
\\
\mathcal{O}_\mathrm{SDW} \!\!&\equiv&\!\! (-1)^j \,
   (n_{j,\uparrow}-n_{j,\downarrow})
,
\\
\mathcal{O}_\mathrm{BCDW} \!\!&\equiv&\!\! (-1)^j 
     (c^\dagger_{j,\uparrow}c_{j+1,\uparrow}^{}
      +c^\dagger_{j,\downarrow}c_{j+1,\downarrow}^{}
      +\mathrm{H.c.})
, \quad\quad
\\
\mathcal{O}_\mathrm{BSDW} \!\!&\equiv&\!\! (-1)^j
     (c^\dagger_{j,\uparrow}c_{j+1,\uparrow}^{}
      -c^\dagger_{j,\downarrow}c_{j+1,\downarrow}^{}
      +\mathrm{H.c.})
. \quad\quad
\end{eqnarray}
\label{eq:order_parameters}%
\end{subequations}
The order parameter of the BCDW state corresponds to the 
   Peierls dimerization operator. 
We note that the BCDW state can be also regarded as the
   $p$-density-wave state,\cite{Nayak}
   as the order parameter of the BCDW state can be written as
   $\sum_j \mathcal{O}_\mathrm{BCDW}\propto \sum_{k,\sigma} \sin(ka)\, 
     c_{k,\sigma}^\dagger \, c_{k+(\pi/a),\sigma}^{}$, where
   $c_{k,\sigma} = N^{-1/2} \sum_j e^{-ikR_j} c_{j,\sigma}$
   with $R_j=ja$ ($a$: the lattice spacing, $N$: the number of sites). 
The BSDW state describes a site-off-diagonal SDW state.\cite{Japaridze}
\begin{figure}[t]
\includegraphics[width=7cm]{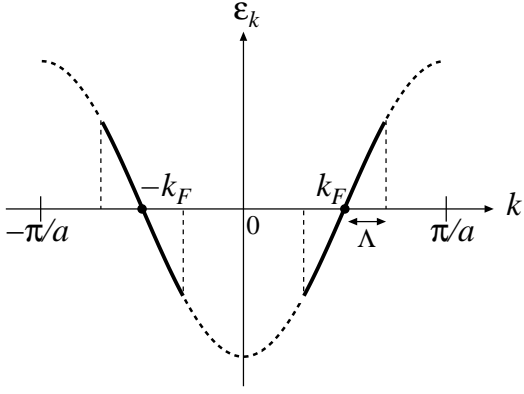}
\caption{
Single-particle energy band.
The annihilation operator of an electron near the Fermi points with
   momentum
   $k\in [-k_F-\Lambda,-k_F+\Lambda]$
   ($k\in[k_F-\Lambda,k_F+\Lambda]$) is denoted $a_{k,-,\sigma}$ 
   ($a_{k,+,\sigma}$), and that of an electron far away from the
   Fermi points is denoted $b_{k,\sigma}$.
}
\label{fig:band}
\end{figure}

\subsection{$g$-ology approach}

The hopping $t$ generates the energy band with dispersion
   $\varepsilon_k=-2t \cos ka$, where
   the Fermi points are at $k=\pm k_F=\pm \pi/2a$ 
   at half filling.
In order to analyze the low-energy physics
   near the Fermi points,
   we introduce a momentum cutoff $\Lambda$ $(0<\Lambda<k_F)$
   and divide the momentum space into the three sectors 
   (Fig.\ \ref{fig:band})
   (i) $k\in R$, (ii) $k\in L$, and (iii) $k \notin (R\cup L)$, 
   where $R=[k_F-\Lambda,k_F+\Lambda]$ and
   $L=[-k_F-\Lambda,-k_F+\Lambda]$.
We then introduce the following fermion operators: 
\begin{equation}
 c_{k,\sigma}
=
\left\{
\begin{array}{cl}
   a_{k,+,\sigma} & \quad  \mbox{for $k\in R$}  \\
   a_{k,-,\sigma} & \quad  \mbox{for $k \in L$} \\
    b_{k,\sigma}  & \quad  \mbox{otherwise}.
\end{array}
\right.
\end{equation}
Electrons near the Fermi points are shuffled by the two-particle
   scattering:
   $H_{\mathrm{int}}=U\sum_j n_{j,\uparrow} \, n_{j,\downarrow}+
    V \sum_j n_{j} \, n_{j+1}$.    
Following the standard $g$-ology approach,\cite{Emery,Solyom}
   we will focus on the scattering processes 
   between electrons near the Fermi points,
   i.e.,
   the scattering processes which involve
   $a_{k,\pm,\sigma}$ only.
The Hamiltonian for such interaction processes is
\begin{eqnarray}
H_\mathrm{int}
&\!\!=\!\!& 
{} +
\frac{g_{1\parallel}}{2L}
\sum_{k_i,p,\sigma}
:\!a_{k_1,p,\sigma}^\dagger \,\,  a_{k_2,-p,\sigma}^{} \,\,
   a_{k_3,-p,\sigma}^\dagger \,\, a_{k_4,p,\sigma}^{} \!:
\nonumber \\ &&
{}+
\frac{g_{1\perp}}{2L}
\sum_{k_i,p,\sigma}
:\!a_{k_1,p,\sigma}^\dagger \,\,  a_{k_2,-p,\sigma}^{} \,\,
   a_{k_3,-p,\overline{\sigma}}^\dagger \,\, 
   a_{k_4,p,\overline{\sigma}}^{} \!:
\nonumber \\ &&
{} +
\frac{g_{2\parallel}}{2L}
\sum_{k_i,p,\sigma}
:\! a_{k_1,p,\sigma}^\dagger \,\, a_{k_2,p,\sigma}^{} \,\,
   a_{k_3,-p,\sigma}^\dagger \,\, a_{k_4,-p,\sigma}^{} \!: 
\nonumber \\ &&
{} +
\frac{g_{2\perp}}{2L}
\sum_{k_i,p,\sigma}
:\! a_{k_1,p,\sigma}^\dagger \,\, a_{k_2,p,\sigma}^{} \,\,
   a_{k_3,-p,\overline{\sigma}}^\dagger \,\, 
   a_{k_4,-p,\overline{\sigma}}^{} \!: 
\nonumber \\ &&
{} +
\frac{g_{3\parallel}}{2L}
\sum_{k_i,p,\sigma}
:\!a_{k_1,p,\sigma}^\dagger \,\, a_{k_2,-p,\sigma}^{} \,\,
   a_{k_3,p,\sigma}^\dagger \,\, a_{k_4,-p,\sigma}^{} \!:
\nonumber \\ &&
{} +
\frac{g_{3\perp}}{2L}
\sum_{k_i,p,\sigma}
:\!a_{k_1,p,\sigma}^\dagger \,\, a_{k_2,-p,\sigma}^{} \,\,
   a_{k_3,p,\overline{\sigma}}^\dagger \,\, 
   a_{k_4,-p,\overline{\sigma}}^{} \!:
\nonumber \\ &&
{} +
\frac{g_{4\parallel}}{2L}
\sum_{k_i,p,\sigma}
:\! a_{k_1,p,\sigma}^\dagger \,\, a_{k_2,p,\sigma}^{} \,\,
    a_{k_3,p,\sigma}^\dagger \,\, a_{k_4,p,\sigma}^{} \!:
\nonumber \\ &&
{} +
\frac{g_{4\perp}}{2L}
\sum_{k_i,p,\sigma}
:\! a_{k_1,p,\sigma}^\dagger \,\, a_{k_2,p,\sigma}^{} \,\,
    a_{k_3,p,\overline{\sigma}}^\dagger \,\, 
    a_{k_4,p,\overline{\sigma}}^{} \!:
,
\nonumber \\
\label{eq:HI_fourier}
\end{eqnarray}
   where $\overline{\sigma}={} \downarrow (\uparrow)$ for
   $\sigma={}\uparrow (\downarrow)$, $L$ is the
   length of the system, and $:\ :$ denotes normal ordering.
The summation over the momentum $k_i$ is taken 
   under the condition of the total momentum being conserved
   (equal to $\pm2\pi/a$ for the umklapp scattering).  
The index $p=+/-$ denotes the right-/left-moving electron.
The coupling constants $g_{1\parallel}$ and $g_{1\perp}$ 
   ($g_{3\parallel}$ and $g_{3\perp}$) denote the matrix elements of
   the backward (umklapp) scattering, while
   $g_{2\parallel}$ and $g_{2\perp}$ ($g_{4\parallel}$ and $g_{4\perp}$)
   denote the matrix element of the forward scattering
   with the different (same) branch $p=\pm$.
The index $\parallel$ ($\perp$) of the coupling constants
   denotes the scattering of
   electrons with same (opposite) spins.

\subsection{Vertex corrections}

In the conventional weak-coupling approach to the
   1D EHM,\cite{Emery,Nakamura}
   one estimates 
   the coupling constants in Eq.\ (\ref{eq:HI_fourier})
   only up to the lowest order in $U$ and $V$:
\begin{subequations}
\begin{eqnarray}
   g_{1\perp}&=&g_{3\perp}=(U-2V)a, \\
   g_{2\perp}&=&g_{4\perp}=(U+2V)a, \\
   g_{1\parallel}&=&g_{3\parallel}=-2Va, \\
   g_{2\parallel}&=&g_{4\parallel}=+2Va.
\end{eqnarray}
\label{eq:g_lowest}%
\end{subequations}
In analyzing the low-energy physics of 
   Eq.\ (\ref{eq:HI_fourier}),
   one then employs the standard $g$-ology approach,\cite{Solyom}
   i.e., the perturbative RG method, and 
   obtains flow equations for the marginal terms 
   in Eq.\ (\ref{eq:HI_fourier}). 
From this RG analysis\cite{Emery,Solyom}
   one finds
   that the $g_{3\perp}$ term generates a gap in the charge excitation
   spectrum
   if $|g_{3\perp}|>-(g_{2\parallel}+g_{2\perp}-g_{1\parallel})$
   and $g_{3\perp}\ne0$,
   whereas the $g_{1\perp}$ term yields a gap in the spin excitation
   spectrum if
   $|g_{1\perp}|>-(g_{2\parallel}-g_{2\perp}-g_{1\parallel})$
   and $g_{1\perp}\ne0$.
Hence, with the lowest-order coupling constants 
   Eq.\ (\ref{eq:g_lowest}), one would conclude that
   the charge (spin) excitations become massless at $U-2V=0$ 
   ($U-2V \ge 0$).
This would mean that, as $U$ increases, 
   both the charge and spin sectors become critical 
   simultaneously at $U=2V$,
   where a direct and continuous CDW-SDW transition takes place.
This analysis is found to be insufficient from the following argument.
The (accidental) simultaneous vanishing
   of $g_{3\perp}$ and $g_{1\perp}$ results from
   the lowest-order estimate in $U$ and $V$
   and there is no symmetry principle that enforces $g_{1\perp}$ and
   $g_{3\perp}$ to vanish simultaneously.
It is possible that
   the higher-order corrections to $g$ lift the degeneracy of
   zeros and change the topology of the phase diagram.
Therefore, in order to analyze the phase diagram at $U\approx 2V$,
   we need to go beyond the lowest-order calculation
   of the coupling constants in the $g$-ology.
In this section, we compute the vertex corrections due to
   virtual processes involving high-energy states \cite{Penc_Mila}
   by integrating out $b_{k,\sigma}$.
This procedure allows us to 
   obtain the effective coupling constants $g$'s
   that include higher-order corrections.

\begin{figure}[t]
 \includegraphics[width=8cm]{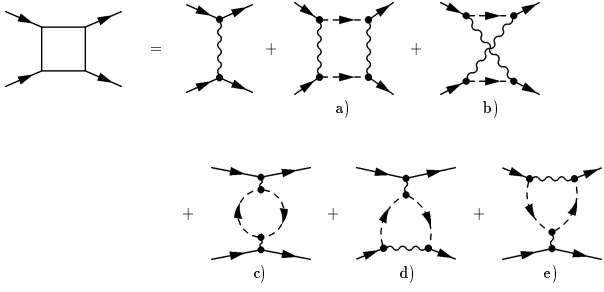}
\caption{
Vertex diagrams with second-order corrections [(a)-(e)].
Solid lines denote electron states
  in the momentum space $k \in R$ or $k \in L$, 
  while the dashed lines denote electron
  states in the other momentum space.
}
\label{fig:vertex}
\end{figure}

The second-order vertex diagrams for the coupling constants
   are shown in Fig.\ \ref{fig:vertex}.
The solid lines denote the low-energy states $a_{k,\pm,\sigma}$,  
   while the dashed lines denote
   high-energy states $b_{k,\sigma}$.
The nonzero contributions from the second-order virtual processes (a)-(e)
are
\begin{subequations}
\begin{eqnarray}
\delta g_{1\perp}^{(a)}
&=&
-\delta g_{3\perp}^{(b)}
=
 -  \frac{U^2}{4\pi t} D_1 a + \frac{V^2}{\pi t} D_2 a , 
\\
\delta g_{1\perp}^{(c)}
&=&
+ \delta g_{3\perp}^{(c)}
=
 + \frac{V(U-2V)}{\pi t} D_1 a , 
\\
\delta g_{2\perp}^{(a)}
&=&
-\delta g_{2\perp}^{(b)}
=
 - \frac{U^2}{4\pi t} D_1 a - \frac{V^2}{\pi t} D_2 a, 
\\
\delta g_{1\parallel}^{(a)}
&=&
 +  \frac{V^2}{\pi t} D_2 a ,  
\\
\delta g_{1\parallel}^{(c)}
&=&
 - \frac{(U-2V)^2+4V^2}{4\pi t} D_1 a 
 - \frac{V^2}{\pi t} D_2 a, 
\\
 \delta g_{2\parallel}^{(a)}
&=&
 - \frac{V^2}{\pi t} D_2 a, 
\\
\delta g_{3\parallel}^{(c)}
&=&
 - \frac{(U-2V)^2+4V^2}{4\pi t} D_1 a 
 + \frac{V^2}{\pi t} D_2 a, 
\end{eqnarray}
\end{subequations}
  where
\begin{subequations}
\begin{eqnarray}
D_1(\Lambda) &\equiv& 
\int_{-\pi/2+a\Lambda}^{\pi/2-a\Lambda} \frac{dk}{\cos k}
,
\\
D_2(\Lambda) &\equiv& 
\int_{-\pi/2+a\Lambda}^{\pi/2-a\Lambda} dk \,\frac{\sin^2 k}{\cos k}
.
\end{eqnarray} 
\end{subequations}
By introducing  $C_1(\Lambda)\equiv 2\ln [\cot (a\Lambda/2)]$ and
    $C_2(\Lambda)\equiv 2\cos a\Lambda $,
    $D_1(\Lambda)$ and $D_2(\Lambda)$ are rewritten as
  $D_1(\Lambda)=C_1(\Lambda)$ and 
  $D_2(\Lambda)=C_1(\Lambda)-C_2(\Lambda)$.
In terms of $C_1$ and $C_2$, the coupling constants 
   with second-order corrections are given by
\begin{subequations}
\begin{eqnarray}
g_{1\perp} \!\! &=& \!\! 
(U-2V)a \left[1
-\frac{C_1}{4\pi t}(U-2V) \right]
-\frac{C_2}{\pi t}V^2a ,
\hspace*{1cm}
\label{eq:g1perp}
\\
g_{1\parallel} \!\! &=& \!\! 
-2Va 
-\frac{C_1}{4\pi t}(U-2V)^2a
-\frac{C_2}{\pi t}V^2a ,
\\
g_{3\perp} \!\!  &=& \!\! 
(U-2V)a \left[1 
+\frac{C_1}{4\pi t}(U+6V)\right]
+\frac{C_2}{\pi t}V^2a ,
\label{eq:g3perp}
\\
g_{3\parallel} \!\!  &=& \!\! 
-2Va 
-\frac{C_1}{4\pi t}(U-2V)^2 a
+\frac{C_2}{\pi t}V^2a ,
\end{eqnarray}
\label{eq:g}%
\end{subequations}
   and $g_{2\parallel} = + 2Va$, $g_{2\perp} = (U+2V)a$,
   $g_{4\parallel} = + 2Va$, and $g_{4\perp} = (U+2V)a$.
Except when $a\Lambda\ll1$, the $C_i$s depend on $\Lambda$ only weakly,
   and we can set $\Lambda=\pi/4$ in the following analysis
   as we are interested in the qualitative feature of the phase
   diagram (different choices will only lead
   to small quantitative changes in phase boundaries).
Incidentally, the logarithmic divergence of $C_1(\Lambda)$ in the
   limit $\Lambda\to0$ leads to the familiar one-loop RG equations.

\subsection{Bosonization}

Having integrated out the high-energy virtual scattering processes,
  we now focus on the low-energy states and
  linearize the dispersion of $a_{k,\pm,\sigma}$ around
  the Fermi points. 
The kinetic-energy term with the linearized dispersion is given by
\begin{eqnarray}
H_{0}
&=&
\sum_{k\in R,\sigma} 
v_F (k-k_F) \, a_{k,+,\sigma}^\dagger \, a_{k,+,\sigma}
\nonumber \\ && {}
+ \sum_{k\in L,\sigma} 
v_F (-k-k_F) \, a_{k,-,\sigma}^\dagger \, a_{k,-,\sigma}
,
\label{eq:H0_linear}
\end{eqnarray}
   where $v_F=2ta$ is the Fermi velocity. 
The field operators of the right- and left-moving electrons 
   are given by
\begin{subequations}
\begin{eqnarray}
\psi_{+,\sigma}(x) &\equiv&
\frac{1}{\sqrt{L}} \sum_{k\in R}
e^{ikx} \, a_{k,+,\sigma}
,
\\
\psi_{-,\sigma}(x) &\equiv&
\frac{1}{\sqrt{L}} \sum_{k\in L}
e^{ikx} \, a_{k,-,\sigma}
.
\end{eqnarray}
\label{eq:psi_fourier_trans}%
\end{subequations}
We apply the Abelian bosonization method and rewrite the
   kinetic-energy term $H_0=\int dx \, \mathcal{H}_0$ in terms of bosonic
   phase fields as (see Appendix \ref{sec:bosonization}) 
\begin{eqnarray}
\mathcal{H}_{0}
&=&
\frac{v_F}{4\pi}
\left[
 \left(2\pi \Pi_{\theta}\right)^2 + \left(\partial_x \theta\right)^2
\right]
\nonumber \\
&& {}
+
\frac{v_F}{4\pi}
\left[
 \left(2\pi \Pi_{\phi}\right)^2 + \left(\partial_x \phi\right)^2
\right]
,
\label{eq:H0}
\end{eqnarray}
   where $\theta$ ($\phi$) is the bosonic field whose spatial
   derivative is proportional to the charge (spin) density,
   $[\theta(x),\phi(y)]=0$.
The operators $\Pi_\theta$ and $\Pi_\phi$ are canonically conjugate
   variables to $\theta$ and $\phi$, respectively, and satisfy
   the conventional commutation relations,
   $[\theta(x),\Pi_\theta(x')]=[\phi(x),\Pi_\phi(x')]=i\delta(x-x')$.
We also introduce chiral bosonic fields
\begin{eqnarray}
\theta_\pm (x)
&\equiv&
\frac{1}{2}
\left[
  \theta(x) \mp 2\pi \int_{-\infty}^x dx' \, \Pi_\theta(x') 
\right]
,
\label{eq:chiral_theta}
\\
\phi_\pm (x)
&\equiv&
\frac{1}{2}
\left[
  \phi(x) \mp 2\pi \int_{-\infty}^x dx' \, \Pi_\phi(x') 
\right]
.
\label{eq:chiral_phi}
\end{eqnarray}
One can easily verify that these chiral fields
   satisfy the commutation relations
  $[\theta_\pm (x),\theta_\pm (x')]=
   [\phi_\pm (x),\phi_\pm (x')]=\pm i (\pi/2) \, \mathrm{sgn}(x-x')$ and
  $[\theta_+ (x),\theta_- (x')]=
   [\phi_+ (x),\phi_- (x')]= i\pi/2$.
In terms of these fields,
   the kinetic-energy density reads
\begin{equation}
\mathcal{H}_{0}
=
\frac{v_F}{2\pi}
\sum_{p=+,-}
\left[
 \left(\partial_x \theta_p\right)^2 + \left(\partial_x \phi_p\right)^2
\right]
.
\end{equation}

To express the electron field operators $\psi_{p,\sigma}$ 
   with the bosonic phase fields,
   we introduce a new set of chiral bosonic fields 
\begin{equation}
\varphi_{p,\uparrow} = \theta_p + \phi_p, \quad
\varphi_{p,\downarrow} = \theta_p - \phi_p, 
\label{eq:varphi}
\end{equation}
   which obey the commutation relations 
\begin{subequations}
\begin{eqnarray}
\left[ \varphi_{\pm,\sigma}(x),\varphi_{\pm,\sigma'}(x') \right]
   &=&
 \pm i \pi \,{\rm sgn}(x-x') \, \delta_{\sigma,\sigma'}
, \quad\quad
\\
\left[\varphi_{+,\sigma}(x),\varphi_{-,\sigma'}(x') \right]
&=&
 i\pi \, \delta_{\sigma,\sigma'}
.
\end{eqnarray}
\label{eq:commutation_varphi}
\end{subequations}
In terms of the phase fields $\varphi_{p,\sigma}$,
   the electron field operators can be written as
\begin{equation}
\psi_{p,\sigma}(x)
  =
   \frac{\eta_\sigma}{\sqrt{2\pi a}}
   \exp\left[ipk_F x+ip \, \varphi_{p,\sigma}(x)\right],
\label{eq:field_op}
\end{equation}
where the Klein factor $\eta_\sigma$ satisfies the anticommutation
   relation $\{\eta_\sigma, \eta_{\sigma'}\}=2\delta_{\sigma,\sigma'}$.
One can verify that the operator defined in Eq.\ (\ref{eq:field_op})
   satisfies the same anticommutation relation as the
   fermion field operator. 
It follows from Eq.\ (\ref{eq:field_op}) that
   the order parameters in Eq.\ (\ref{eq:order_parameters})
    are rewritten as
\begin{subequations}
\begin{eqnarray}
\mathcal{O}_\mathrm{SDW}(x)
&\propto&
\cos\theta(x) \, \sin \phi(x),
\label{eq:ordersdw}
\\
\mathcal{O}_\mathrm{CDW}(x)
&\propto&
\sin\theta(x) \, \cos \phi(x),
\label{eq:ordercdw}
\\
\mathcal{O}_\mathrm{BCDW}(x)
&\propto&
\cos\theta(x) \, \cos \phi(x),
\\
\mathcal{O}_\mathrm{BSDW}(x)
&\propto&
\sin\theta(x) \, \sin \phi(x).
\label{eq:orderbsdw}
\end{eqnarray}
\label{eq:orderparam}
\end{subequations}

The interaction part of the Hamiltonian  
   $H_\mathrm{int}$, Eq.\ (\ref{eq:HI_fourier}), can be also
   expressed in terms of the boson fields $\theta_\pm$ and $\phi_\pm$.
It has been suggested that, besides the marginal operators,
   operators with  higher scaling dimensions can play an important role
   in the first-order CDW-SDW transition \cite{Cannon,Voit}
   which is known to occur in the strong-coupling region
   of the 1D EHM. \cite{Emery,Bari,Hirsch,Dongen}
We thus include all the terms of scaling dimension 4
   $[=2\, (\mbox{charge sector}) + 2\, (\mbox{spin sector})]$.
We also note that there are some complications and subtleties in
   bosonizing the off-site interaction term,
   i.e., the nearest-neighbor interaction term $V$
   (see Appendix \ref{sec:bosonization} for detail).
We obtain the bosonized Hamiltonian density,
\begin{eqnarray}
\mathcal{H} &=&
\frac{1}{2\pi}\sum_{p=+,-}
 \left[v_\rho(\partial_x\theta_p)^2
      +v_\sigma(\partial_x\phi_p)^2\right]
\nonumber \\ && {}
+ \frac{g_\rho}{2\pi^2}
     \left(\partial_x \theta_+ \right)
     \left(\partial_x \theta_- \right)
- \frac{g_\sigma}{2\pi^2}
     \left(\partial_x \phi_+ \right) 
     \left(\partial_x \phi_- \right)
\nonumber \\ && {}
-\frac{g_c}{2\pi^2 a^2} \, \cos 2 \theta 
+\frac{g_s}{2\pi^2 a^2} \, \cos 2 \phi
\nonumber \\ && {}
-\frac{g_{cs}}{2\pi^2 a^2} \, \cos 2\theta  \,  \cos 2\phi
\nonumber \\ && {}
-\frac{g_{\rho s}}{2 \pi^2}
     \left(\partial_x \theta_+ \right) 
     \left(\partial_x \theta_- \right) \, \cos 2\phi
\nonumber \\ && {}
+\frac{g_{c\sigma}}{2 \pi^2}
     \left(\partial_x \phi_+\right) 
     \left(\partial_x \phi_- \right) \, \cos 2\theta 
\nonumber \\ && {}
+\frac{g_{\rho \sigma}}{2\pi^2}  \, a^2 
     \left(\partial_x \theta_+ \right) 
     \left(\partial_x \theta_- \right) 
     \left(\partial_x \phi_+ \right) 
     \left(\partial_x \phi_- \right) 
.
\label{eq:Hamiltonian}
\end{eqnarray}
The renormalized velocities are
   $v_\rho=2ta+(g_{4\parallel}+g_{4\perp}-g_{1\parallel})/2\pi$ and
   $v_\sigma=2ta+(g_{4\parallel}-g_{4\perp}-g_{1\parallel})/2\pi$.
The marginal terms with the couplings $g_{\rho}$ and $g_{c}$ 
   ($g_{\sigma}$ and $g_{s}$)
   determine low-energy properties of the charge (spin)
   modes, \cite{Emery,Solyom}
   where 
   $g_\rho=g_{2\perp}+g_{2\parallel}-g_{1\parallel}$,
   $g_c=g_{3\perp}$,
   $g_\sigma=g_{2\perp}-g_{2\parallel}+g_{1\parallel}$,
   and $g_s=g_{1\perp}$.
The $g_{cs}$, $g_{\rho s}$, $g_{c\sigma}$, and $g_{\rho\sigma}$ terms
   with scaling dimension 4
   couple the spin and charge degrees of freedom.
The $g_{cs}$ coupling comes from the 
   umklapp scattering $g_{3\parallel}$.
The $g_{\rho s}$ ($g_{\rho\sigma}$) coupling is generated 
   from the backward scattering of antiparallel- (parallel-) spin
   electrons while the $g_{c\sigma}$ coupling is generated
   from the umklapp scattering of electrons with antiparallel spins
   (see Appendix \ref{sec:bosonization}).
These coupling constants are given by
   $g_{cs}=g_{\rho s}=g_{c\sigma}=g_{\rho\sigma}=-2Va$ to lowest order
   in $V$.
Cannon and Fradkin examined the effect of the $g_{cs}$ term
   and argued that it plays a crucial role in 
   the first-order CDW-SDW transition.\cite{Cannon}
Voit included the $g_{\rho s}$ and $g_{c\sigma}$ terms, as well as the
   $g_{cs}$ term, in the perturbative RG analysis of the coupling
   constants, but did not consider the $g_{\rho\sigma}$ term.\cite{Voit}
Here we note that
   it is important to keep the $g_{\rho\sigma}$ term as well, 
   since the global SU(2) symmetry in the spin sector 
   is guaranteed only when $g_\sigma=g_s$,
   $g_{cs}=g_{c\sigma}$, and $g_{\rho s}=g_{\rho\sigma}$.

\subsection{Renormalization-group equations}

We perform a perturbative RG calculation to examine the low-energy
   properties of the 1D EHM in the weak-coupling regime, taking into
   account quantum fluctuations of the phase fields.
The operator product expansion (OPE) technique allows us
   to systematically handle the higher-order terms in the bosonized
   Hamiltonian (\ref{eq:Hamiltonian}).
The one-loop RG equations that describe changes in the coupling
   constants during the scaling of the short-distance cutoff
   ($a\to ae^{dl}$) are  given by
   (see Appendix \ref{sec:rg} for their derivation)
\begin{eqnarray}
\frac{d}{dl} G_\rho 
  &=& {}
    + 2 \, G_c^2 + G_{cs}^2 +  G_s \, G_{\rho s} ,
\label{eq:Grho}
\\
\frac{d}{dl} G_c 
  &=& {}
    + 2 \, G_\rho\, G_c - G_s \, G_{cs} - G_{cs} \, G_{\rho s} ,
\label{eq:Gc}
\\
\frac{d}{dl} G_s 
  &=& {}
    - 2 \, G_s^2 - G_c \, G_{cs} - G_{cs}^2 ,
\label{eq:Gs}
\\
\frac{d}{dl} G_{cs} 
  &=& {}
    - 2 \, G_{cs} + 2 \, G_\rho \, G_{cs} - 4 \, G_s \, G_{cs}
\nonumber \\ && {}
    - 2 \, G_c\, G_{s}
    - 2 \, G_c\, G_{\rho s}
    - 4 \, G_{cs} \, G_{\rho s} , \quad\quad
\label{eq:Gcs}
\\
\frac{d}{dl} G_{\rho s} 
  &=& {}
    - 2 \, G_{\rho s} + 2 \, G_\rho \, G_s
\nonumber \\ && {}
    - 4 \, G_c \, G_{cs} - 4 \, G_{cs}^2
    - 4 \, G_s \, G_{\rho s} ,
\label{eq:Grhos}
\end{eqnarray}
   where $G_\nu$ are dimensionless coupling constants with
   the initial values $G_\nu (0)=g_\nu /(4\pi ta)$.
The number of the independent coupling constants is five, 
   since the SU(2) spin symmetry guarantees the relations 
   $G_\sigma=G_s$,
   $G_{c\sigma}=G_{cs}$, and $G_{\rho\sigma}=G_{\rho s}$ to hold
   in the scaling procedure.
From these scaling equations, one finds that
   the $G_\rho$, $G_c$, and $G_s$ terms are marginal
   (the scaling dimension=2),\cite{Cardy,Gogolin}
   while the $G_{cs}$ and $G_{\rho s}$ terms are irrelevant operators
   of the scaling dimension 4.

\section{Phase Diagram of the Half-Filled Extended Hubbard Model}
\label{sec:phase_diagram}

\subsection{Bond-charge-density-wave state}\label{sec:BCDW}

In this section,
   we show that the BCDW phase exists 
   in between the CDW and SDW phases
   in the weak-coupling region of the 1D EHM.

First we focus on the weak-coupling limit $U,V\ll t$,
   where we can neglect the irrelevant terms of scaling dimension 4
   and restrict ourselves to the marginal terms $\propto$
   $g_\rho$, $g_\sigma$, $g_c$, and $g_s$.
Effects of the irrelevant terms
   are discussed later in this section.
Within this approximation, the Hamiltonian reduces to two decoupled
   sine-Gordon models, and 
   we can analyze the properties of the spin and 
   charge modes, separately.
The one-loop RG equations for these coupling constants
   are given by Eqs.\ (\ref{eq:Grho})--(\ref{eq:Gs})
   with $G_{cs}=G_{\rho s}=0$: 
\begin{eqnarray}
\frac{d}{dl} G_\rho(l) &=& 2 \, G_{c}^2(l),
\label{eq:Grho1}
\\ 
\frac{d}{dl} G_{c}(l) &=& 2 \, G_\rho(l) \, G_{c}(l),
\label{eq:Gc1}
\\
\frac{d}{dl} G_{s}(l)  &=& - 2 \, G_{s}^2(l).
\label{eq:Gs1}
\end{eqnarray}

The spin excitations are controlled by the $G_s$ coupling,
   which is marginally relevant (marginally irrelevant)
   when $G_s<0$ ($G_s>0$).
If $g_{s}<0$, then $|G_{s}(l)|$ increases with increasing $l$.
In this case the phase field $\phi$ is locked at $\phi=0$ mod $\pi$
   to gain the energy [see Eq.\ (\ref{eq:Hamiltonian})],
   and consequently the spin excitations have a gap.
On the other hand, if $g_{s}>0$, then $|G_{s}(l)|$ decreases to zero
   as $l$ increases, and the $\phi$ field becomes a free field;
   the spin sector has massless excitations.
The approach of $G_s$ to zero is very slow ($\sim1/l$), and the
   $\phi$ field has a strong tendency to be near $\phi=\pi/2$ mod $\pi$.
Although it eventually fails to lock the phase $\phi$,
   the marginally irrelevant coupling still has an impact on
   low-energy properties by giving rise
   to logarithmic corrections to
   correlation functions.\cite{Giamarchi1989} 

The charge sector is governed by the two couplings $G_c$ and $G_\rho$, 
   whose RG flow diagram is of the Kosterlitz-Thouless type.
Since $g_\rho=(U+6V)a >0$, $G_c$ is a relevant coupling and always
   flows to strong-coupling regime, unless $g_{c}=0$.
This means that $G_{c}(l)$ has two
   strong-coupling fixed points,
   $G_{c}(l)\to \infty$ and $G_{c}(l)\to -\infty$,
   depending on its initial value $g_{c}>0$ and $g_{c}<0$.
As seen from Eq.\ (\ref{eq:Hamiltonian}), the relevant $g_c$ 
   with positive (negative) sign implies the 
   phase locking of $\theta$ at the position 
   $\theta=0$ $(\pi/2)$ mod $\pi$.

\begin{figure}[b]
\includegraphics[width=7.cm]{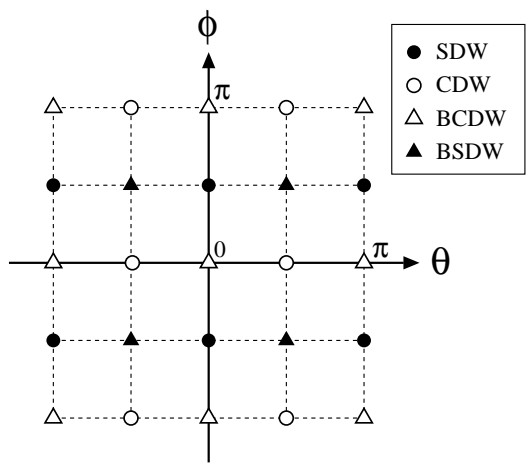}
\caption{
Positions of locked phase fields $\theta$ and $\phi$
   in the SDW, CDW, BCDW, and BSDW states. 
}
\label{fig:potmin}
\end{figure}

From the above standard arguments, 
   the ground state can be identified by simply looking at the
   initial value of the coupling constants
   $g_{c}$ and $g_{s}$.
The ground state is classified into four cases as summarized
   in Table \ref{table:phase-locking}, and
   the positions of locked phases $(\theta,\phi)$ for respective cases 
   are shown in Fig.\ \ref{fig:potmin}. 

(i) $g_{s}<0$ and $g_{c}<0$: The phase fields are locked at 
   $(\theta,\phi)
   =\biglb( (\pi/2)+\pi I_1,\pi I_2\bigrb)$,
   where $I_1$ and $I_2$ are integers.
In this case, among the order parameters in Eqs.\ (\ref{eq:orderparam}),
   only the CDW order parameter has a finite expectation value, and
   the ground state is found to be the CDW state.
Both charge and spin excitations are gapped.

(ii) $g_s<0$ and $g_c>0$: The phase fields are locked at
   $(\theta,\phi)=(\pi I_1,\pi I_2)$.
The nonvanishing order parameter is then $\mathcal{O}_\mathrm{BCDW}$,
   and the ground state is the BCDW state.
Both charge and spin excitations are gapped.

(iii) $g_s>0$ and $g_c<0$: The field $\theta$ is locked at
   $\theta=(\pi/2)+\pi I_1$, and the field $\phi$ tends to be
   around $\phi=(\pi/2)+\pi I_2$ although it is not locked in the
   low-energy limit.
In this case the dominant correlation is that of the BSDW state.
The charge excitations are gapped whereas the spin excitations
   are gapless.

(iv) $g_s>0$ and $g_c>0$: The field $\theta$ is locked at
   $\theta=\pi I_1$, whereas the field $\phi$ tends to be near
   $\phi=(\pi/2)+\pi I_2$.
The dominant correlation is the SDW order.
The charge excitations are gapped while the spin excitations are
   gapless.

\begin{table}
\caption{
Possible ground-state phases and positions of (quasi)
   locked phase fields
   determined only from the marginal terms 
   in Eq.\ (\protect{\ref{eq:Hamiltonian}}).
}
\label{table:phase-locking}
\begin{ruledtabular}
\begin{tabular}{lcc}
Phase &  $(\theta,\phi)$    &  $(g_c,g_s)$   \\
\hline
SDW  & $(0,\pm \pi/2),
        (\pi,\pm \pi/2)$
     & $(+,+)$ \\
CDW  & $(\pm \pi/2,0), 
        (\pm \pi/2,\pi)$
     & $(-,-)$  \\
BCDW & $(0,0),(\pi,\pi),
        (0,\pi),(\pi,0)$              
     & $(+,-)$    \\
BSDW & $(\pm \pi/2, 
        \pm \pi/2)$ 
     & $(-,+)$
\end{tabular}
\end{ruledtabular}
\end{table}

Combining the results of Table \ref{table:phase-locking} and
  the coupling constants Eqs.\ (\ref{eq:g1perp}) and (\ref{eq:g3perp}),
  we obtain the ground-state phase diagram
  of the 1D EHM in the weak-coupling limit.
For $U$ larger than $2V$ such that $g_c>0$ and $g_s>0$,
   we have the SDW phase, while
   for $U$ sufficiently smaller than $2V$ ($g_c<0$ and $g_s<0$)
   we have the CDW phase.
At $U=2V$, we see from Eqs.\ (\ref{eq:g1perp}) and (\ref{eq:g3perp})
   that $g_s(=g_{1\perp})<0$ and $g_c(=g_{3\perp})>0$  due to
   the $C_2$ term.
This implies that a new phase 
   different from the CDW and SDW states
   appears for $U\approx 2V$.
From Table \ref{table:phase-locking}, 
   we identify the new phase with the BCDW phase.
Within the approximation we employ here, the phase boundary between
   the BCDW phase and the CDW (SDW) phase is located at $g_c=0$
   ($g_s=0$).
In this phase diagram,
   the charge excitations are gapful except on the CDW-BCDW transition
   line,
   while the spin excitations are gapless in the SDW phase and on the
   SDW-BCDW transition line.
From Eqs.\ (\ref{eq:Grho1})--(\ref{eq:Gs1}), we can estimate
  the charge gap $\Delta_c$ and the spin gap
   $\Delta_s$ as 
\begin{equation}
  \Delta_c\approx t \left(\frac{|g_c|}{ta}\right)^{2\pi t a/g_\rho},
 \quad
  \Delta_s\approx  t\exp\left( \frac{2\pi t a}{g_s}\right)
\end{equation}
for $|g_c|\ll g_\rho \ll  ta$ and
   $0<-g_s\ll ta$, respectively.

Next we examine effects of the parallel-spin umklapp scattering
   $g_{cs}$ on the BCDW state.
We consider the situation very close to the CDW-BCDW
   transition by assuming
   $g_c\approx0$ and $g_s<0$, i.e.,
   $U-2V = -C_2V^2/\pi t + O(V^3/t^2)$.
In this case the spin gap is formed first as the energy scale is
   lowered.
For energies below the spin gap,
   we can replace $\cos2\phi$ with its average
   $\langle\cos2\phi\rangle\approx (\Delta_s/t)^2$.
This means that the coupling constant $g_c$ 
   is modified as
\begin{equation}
g_{c}^* = g_c + g_{cs} \langle\cos2\phi\rangle.
\label{eq:gc*}
\end{equation}
Thus we find that the BCDW state,
   which is realized for $g_c^*>0$, becomes
   less favorable due to the 
   $g_{cs}(<0)$ term.
We note, however, that the CDW-BCDW boundary does not move across
   the $U=2V$ line because
   $|g_{cs}\langle\cos2\phi\rangle|\approx 2Va\exp[-c(t/V)^2]$
   is much smaller than the $C_2$ term 
   in Eq.\ (\ref{eq:g3perp}) for $V\ll t$,
   where $c$ is a positive constant.
A similar argument applies to the region near the SDW-BCDW
   transition.
Suppose that $U-2V=+C_2V^2/\pi t + O(V^3/t^2)$
   where $g_s\approx0$ and $g_c>0$.
In this case, as the energy scale is lowered, the charge gap
   opens first and the $\theta$ field is pinned at $\theta=0$
   mod $\pi$.
Below the charge-gap energy scale, the $\phi$ field is subject to
   the pinning potential $g_s^*\cos2\phi$ with
\begin{equation}
g_{s}^* = g_s - g_{cs} \langle\cos2\theta\rangle,
\label{eq:gs*}
\end{equation}
  where $\langle\cos2\theta\rangle\approx (\Delta_c/t)^{2(1-G_\rho)}$.
Thus the BCDW phase, which is now realized 
   for $g_s^*<0$, also becomes less favorable by the 
   $-g_{cs}\langle\cos2\theta\rangle(>0)$ term.
Again the phase boundary is not moved beyond the $U=2V$ line since
   $|g_{cs}\langle\cos2\theta\rangle|\approx 2Va(c'V/t)^{\pi t/V}$
   is much smaller than the $C_2$ term
   in Eq.\ (\ref{eq:g1perp}), where $c'$ is a constant of order
   1. 
Therefore we conclude that 
    the BCDW phase is robust against the $g_{cs}$ term
    in the weak-coupling limit.
The analysis in this section establishes the existence of the
   BCDW phase near $U\approx2V$ for $0<U,V\ll t$.

\subsection{First-order SDW-CDW transition}\label{sec:SDW-CDW_trans}

In this section, we discuss how the BCDW phase becomes unstable
   at strong coupling and how the 
   two continuous transitions change into the 
   first-order SDW-CDW transition.

To our knowledge,
Cannon and Fradkin were the first to argue that 
   the $g_{3\parallel}$ term (describing the umklapp scattering of
   parallel-spin electrons),
   which is conventionally ignored due to its
   large scaling dimension,
   can become relevant at large $U$ and $V$
   and cause the first-order CDW-SDW transition. \cite{Cannon}
To get an insight into the effect of the 
   $g_{cs}$ term in the relevant case, 
   we perform a semiclassical analysis:
   we neglect spatial variations
   of the fields in Eq.\ (\ref{eq:Hamiltonian})
   and focus on the potential,
\begin{equation}
 V(\theta,\phi)=
    - g_c\cos2\theta + g_s\cos 2\phi
    - g_{cs} \cos2\theta \, \cos 2\phi,
\label{eq:V}
\end{equation}
   where $g_{cs}=g_{3\parallel}<0$.
The order parameters of the SDW, CDW, BCDW, and BSDW states
   take maximum amplitudes when the fields
   $\theta$ and $\phi$ are pinned at 
   $(\theta,\phi)
   =\biglb(\pi I_1,(\pi/2)+\pi I_2\bigrb)$,
   $\biglb( (\pi/2)+\pi I_1,\pi I_2\bigrb)$,
   $(\pi I_1,\pi I_2)$, and 
   $\biglb( (\pi/2)+\pi I_1,(\pi/2) + \pi I_2\bigrb)$,
   respectively, where $I_1$ and $I_2$ are integers.
The potential energy in these states is obtained
   by inserting these pinned fields
   into Eq.\ (\ref{eq:V}), e.g,
   $V_{\mathrm{SDW}}=V\biglb(\pi I_1,(\pi/2)+\pi I_2\bigrb)$, yielding
\begin{subequations}
\begin{eqnarray}
V_\mathrm{SDW}  &=& - g_c-g_s-|g_{cs}|  , \\
V_\mathrm{CDW}  &=& + g_c+g_s-|g_{cs}|  , \\
V_\mathrm{BCDW} &=& - g_c+g_s+|g_{cs}|  , \\
V_\mathrm{BSDW} &=& + g_c-g_s+|g_{cs}|  .
\end{eqnarray}
\end{subequations}
\begin{figure}[t]
\includegraphics[width=7cm]{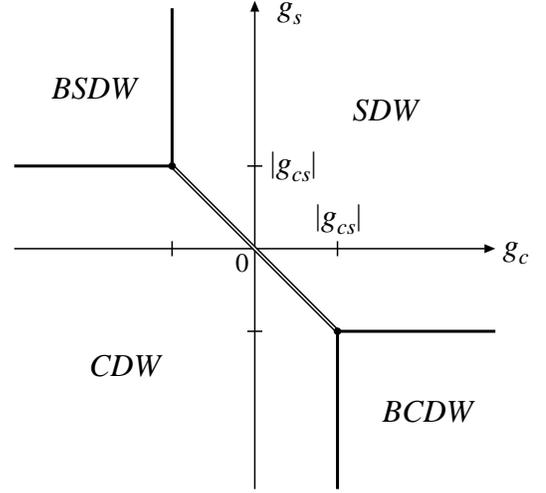}
\caption{
Phase diagram obtained by minimizing the potential $V(\theta,\phi)$
   for $g_{cs}<0$.
The double line denotes the first-order transition, while the single
   line denotes the second-order transition.
Bicritical points are at
   $(g_c,g_s)=(\pm |g_{cs}|,\mp |g_{cs}|)$.
}
\label{fig:classic}
\end{figure}%
We find that
   the $g_{cs}$ term stabilizes
   the SDW and CDW states
   while it works against the BCDW and BSDW states.
Comparing these energies, we obtain the phase diagram in the
   $g_c$-$g_s$ plane at a fixed $g_{cs}$ (Fig.\ \ref{fig:classic}).
In the presence of the $g_{cs}$ term, the direct
   CDW-SDW transition line appears 
   in this phase diagram.

\begin{figure}[t]
\includegraphics[width=7cm]{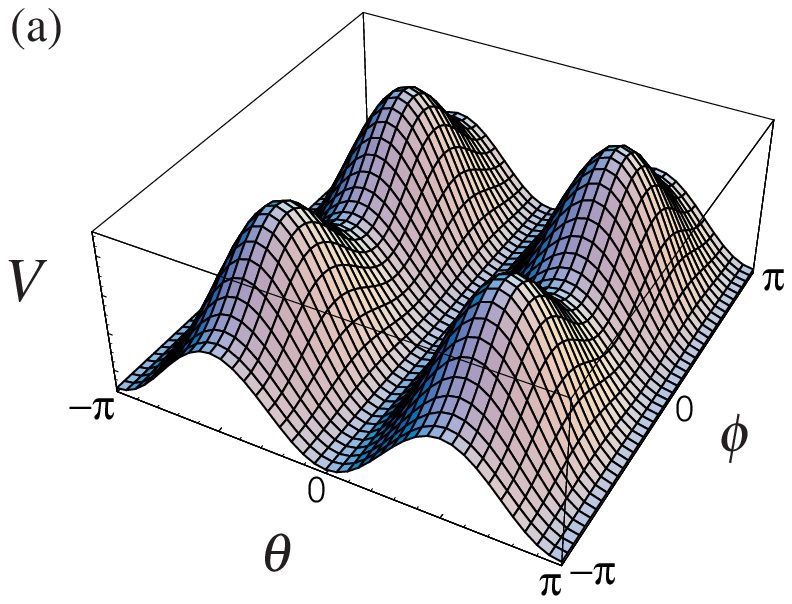}
\includegraphics[width=7cm]{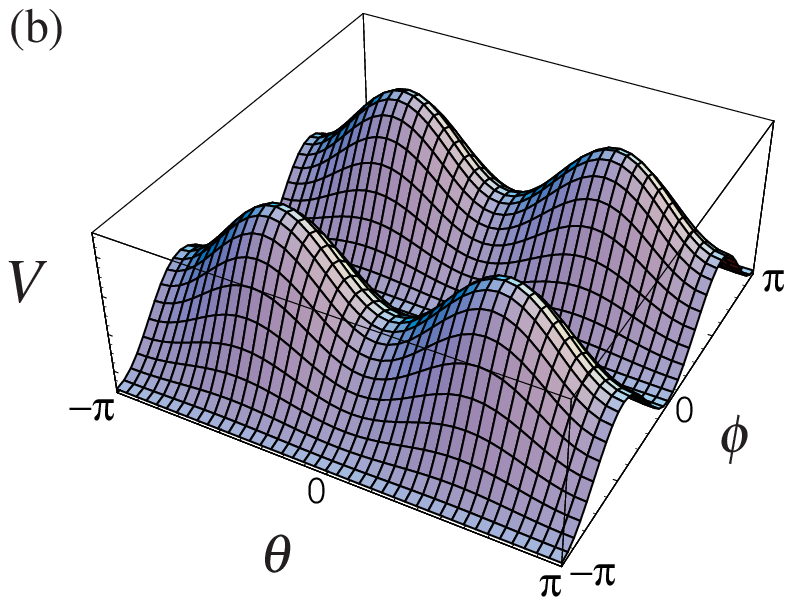}
\includegraphics[width=7cm]{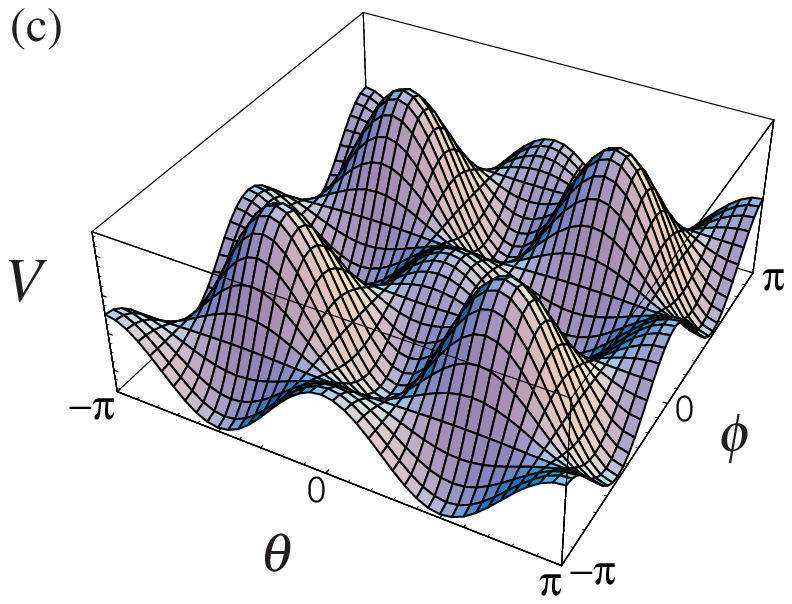}
\caption{
The potential $V(\theta,\phi)$ on 
   the SDW-BCDW (a), BCDW-CDW (b), and CDW-SDW (c) transition lines.
}
\label{fig:v}
\end{figure}

We now discuss the nature of the phase transitions.
The potential $V(\theta,\phi)$ on various
   transition lines is shown in Fig.\ \ref{fig:v}.
On the boundary between the SDW and BCDW phases, which
   is located at $g_s=-|g_{cs}|$ and $g_c>|g_{cs}|$, the potential 
   takes the form $V(\theta,\phi)=
    -g_c \cos 2\theta + g_s \cos 2\phi (1-\cos 2\theta)$
   [Fig.\ \ref{fig:v}(a)],
   which pins the $\theta$ field at $\theta=\pi I_1$ and leaves
   the $\phi$ field completely free.
We thus find that 
   the SDW-BCDW transition is continuous, i.e.,
   the SDW and BCDW phases coexist
   without potential barrier on the phase boundary.
On the boundary between the BCDW and CDW phases,
   located at $g_c=|g_{cs}|$ and $g_s<-|g_{cs}|$,
   the potential now
   takes the form $V(\theta,\phi)=
    -g_c \cos 2\theta (1-\cos 2\phi)+ g_s \cos 2\phi $
   [Fig.\ \ref{fig:v}(b)].
The potential locks the $\phi$ field at $\phi=\pi I_2$, where 
   it has no effect on the $\theta$ field.
Thus, we find that the CDW-BCDW transition is also continuous. 
From similar considerations, we find that 
   the SDW-BSDW and BSDW-CDW transitions are continuous as well.
In Fig.\ \ref{fig:classic}, the phase boundaries of 
   continuous transitions are shown by the solid lines. 
On the contrary, the phase  
   boundary shown by the double line in Fig.\ \ref{fig:classic}
   is of different nature from the others.
The potential $V(\theta,\phi)$ on the double line is 
   shown in Fig.\ \ref{fig:v}(c), where
   the potential minima  
   are given by the isolated points 
   $(\theta,\phi)
     =\biglb(\pi I_1,(\pi/2) + \pi I_2\bigrb)$
   and $\biglb((\pi/2)+\pi I_1, \pi I_2\bigrb)$.
These minima correspond to the SDW state
   and the CDW state,
   see Fig.\ \ref{fig:potmin}.
The point to note is that
   there is a finite potential barrier of height 
   $\min(|g_{cs}|,2|g_{cs}| - 2|g_c|)$
   between the corresponding minima for the SDW and CDW phases.
Hence we conclude that 
   the CDW-SDW transition is first order when $g_{cs}$ is relevant.

From the above arguments, we find that
   strong umklapp scattering 
   of the parallel-spin electrons destabilizes the BCDW and 
   BSDW states and
   gives rise to bicritical points $(g_c,g_s)=\pm (g_{cs},-g_{cs})$
   where the two continuous-transition lines merge
   into the CDW-SDW first-order transition line.
Let us take a closer look at these bicritical points.
Taking into account the fact that $g_c>0$ and $g_s<0$
   for $U\approx 2V$ in the original EHM,
   we will focus on the bicritical point at
   $(g_c,g_s)=(|g_{cs}|,-|g_{cs}|)$.
The effective potential at the bicritical point takes the form
\begin{figure}[t]
\includegraphics[width=7cm]{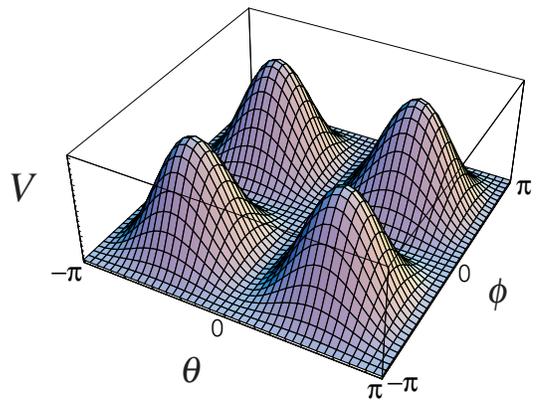}
\caption{
The potential $V(\theta,\phi)$ on 
   the bicritical point $(g_c,g_s)=(|g_{cs}|,-|g_{cs}|)$.
The potential minima are
   the lines $\theta=\pi I_1$ and $\phi= \pi I_2$.
}
\label{fig:v4}
\end{figure}%
\begin{equation}
   V(\theta,\phi)=-g(\cos2\theta+\cos2\phi-\cos2\theta\cos2\phi),
\end{equation}
   which is shown in Fig.\ \ref{fig:v4}.
This potential has an interesting feature that its potential minima are
   not isolated points but the crossing lines $\theta=\pi m$ or
   $\phi=\pi n$ ($m$, $n$: integer).
On these lines either $\theta$ or $\phi$ becomes a free field;
   the theory has more freedom than a single free bosonic field, but
   less than two free bosonic fields.
We thus expect that the theory of the bicritical point should have a
   central charge larger than 1 but smaller than 2.
Detailed analysis of the critical theory is left for a future study.
We note that when $g_{cs}=0$ 
   the first-order CDW-SDW transition line collapses into
   a tetracritical point, $(g_c,g_s)=(0,0)$,
   and the
   phase boundaries in Fig.\ \ref{fig:classic}
   reduce to the lines $g_c=0$ and $g_s=0$
   where all the transitions are continuous.

Fabrizio \textit{et al}.\ \cite{Fabrizio} and Bajnok
   \textit{et al}.\ \cite{Bajnok}
   discussed effects of higher-frequency terms, 
   such as $\sin 3\theta$ and $\cos4\theta$, which are generated 
   through the renormalization-group transformation.
From the semiclassical arguments,
   it can be seen that these terms can also change a second-order
   transition to a first-order transition.\cite{Bajnok}
In fact, it was argued that these higher-frequency terms make the
   SDW-CDW transition first order in the strong-coupling regime of the 
   1D EHM.\cite{Fabrizio}
However,
   we have shown that the SDW-CDW first-order transition can occur
   simply due to the $g_{cs}$ term which is
   the leading irrelevant term in this system.
Since the higher-frequency terms are even less relevant than the
   $g_{cs}$ term, we expect that the $g_{cs}$ term should play a
   dominant role in the first-order transition in the 1D EHM.

\begin{figure}[t]
\includegraphics[width=7cm]{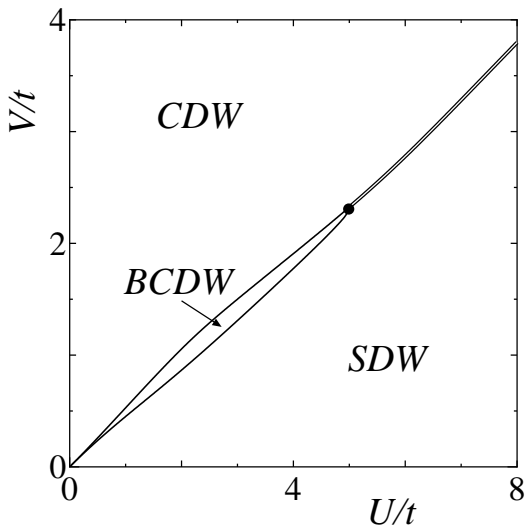}
\caption{
Phase diagram of the half-filled 1D extended Hubbard model.
The double line denotes the first-order transition,
  while the single lines denote the second-order transitions.
The bicritical point is at $(U_c,V_c)\approx (5.0t, 2.3t)$.
}
\label{fig:phase2}
\end{figure}

\subsection{Global ground-state phase diagram}

To obtain the global phase diagram of the 1D EHM,
   we have numerically solved the scaling equations
   (\ref{eq:Grho})--(\ref{eq:Grhos}).
We find out which phase is realized by looking at which one of the
   couplings $G_c$, $G_s$, and $G_{cs}$ becomes relevant first,
   as we have discussed 
   in Secs.\ \ref{sec:BCDW} and \ref{sec:SDW-CDW_trans}.
First, if $|G_c|$ grows with increasing $l$ and reaches, say, 
   1 first among the three couplings,
   then we stop the integration and compute
   $G_s^*=G_s-G_{cs} \, \mathrm{sgn}(G_c)$.
Since the charge fluctuations are suppressed below this energy scale,
   we are left with Eq.\ (\ref{eq:Gs1}), 
   where $G_s$ is replaced by $G_s^*$.
We immediately see from Table \ref{table:phase-locking}
   that a positive (negative) $G_s^*$ leads to the
   SDW (BCDW) state for $G_c>0$ and the BSDW (CDW) state
   for $G_c<0$. 
Second, if $|G_s|$ becomes 1 first,
   or more precisely, if $G_s$ reaches $-1$ first, 
   then we are left with Eqs.\ (\ref{eq:Grho1}) and 
   (\ref{eq:Gc1}),  where $G_\rho$  and $G_c$ are replaced by
   $G_{\rho}^*=G_\rho - G_{\rho s}$ and 
   $G_c^*=G_c+G_{cs}$, respectively.
We see that a positive (negative) $G_c^*$    
   leads to the BCDW (CDW) state.
Finally, when $|G_{cs}|$ reaches 1 first, we stop the calculation and
   compare $G_c$ and $G_s$.
Since both charge and spin fluctuations are already suppressed by the
   $G_{cs}\cos2\theta\cos2\phi$ potential, we can deduce the phase
   from the semiclassical argument.
From Fig.\ \ref{fig:classic} we see that we have the SDW state for
   $G_s>-G_c$ and the CDW state for $G_s<-G_c$.
Here we note that
  in the SDW state the pinning potential to the $\phi$ field is
   marginally irrelevant and 
   thus the spin sector should become gapless.

The phase diagram obtained in this manner is shown in 
   Fig.\ \ref{fig:phase2}.
The single lines denote continuous transitions, and
   the double line denotes the first-order transition.
In the weak-coupling limit, the BCDW phase appears at $U\approx 2V$ and
   the successive continuous transitions between the SDW, BCDW, and
   CDW states occur as $V/U$ increases.
When $U$ and $V$ increase along the line $U\approx 2V$,
   the BCDW phase first
   expands and then shrinks up to the bicritical point
   $(U_c,V_c)\approx(5.0t, 2.3t)$ where the two continuous-transition
   lines meet.
Beyond this point the BCDW phase disappears and we have the direct
   first-order transition between the CDW and  SDW phases.
The phase diagram (Fig.\ \ref{fig:phase2}) is similar to the ones
   obtained by using more sophisticated numerical
   methods.\cite{Nakamura,Sengupta}
We note that the position of the first-order transition line 
   in Fig.\ \ref{fig:phase2} is not reliable quantitatively 
   as we have used the perturbative RG equations.
The recent Monte Carlo calculation\cite{Sengupta} gives the most
   reliable estimate for the position of the bicritical point,
   $(U_c,V_c)\approx\biglb((4.7\pm0.1)t, (2.51\pm0.04)t\bigrb)$, which
   agrees with our estimate in Fig.\ \ref{fig:phase2} within 10\%.
The semiquantitative agreement gives us confidence that our approach,
   semiclassical analysis of the low-energy effective Hamiltonian
   derived with use of the perturbative RG, is reliable even in the
   strong-coupling regime near the multicritical point.

\section{Effect of staggered site potential}\label{sec:staggered}

In this section, we examine effects of
   alternating on-site modulation of the chemical potential, i.e., 
   the staggered site potential, in the half-filled 1D EHM.
The Hamiltonian to be considered is given by $H'=H+H_\Delta$ with 
   $H$ defined in Eq.\ (\ref{eq:H1D}) and
\begin{equation}
H_\Delta =  \Delta \sum_{j,\sigma} (-1)^j \,  n_{j,\sigma}.
\label{eq:H_Delta}
\end{equation}
The model is called the ionic Hubbard model if $V=0$.
When $U=V=0$, the system is a trivial band insulator, since 
   the $\Delta$ term induces a gap $2|\Delta|$ at $k= \pm \pi/2$ in the 
   single-particle spectrum and the lower band is fully filled.
For many years effects of the \textit{on-site} repulsive interaction 
   $U$ on the band insulator have been investigated intensively
   \cite{MJRice,Nagaosa,Girlando,Egami,Fabrizio,%
        Tsuchiizu_JPSJ,Takada,Qin,Brune,YZZhang,Anusooya-Pati,Wilkens,%
        Torio,Refolio,Caprara,Gupta,Pozgajcic,%
        Resta,Gidopoulos,Manmana}
   from both numerical and analytical approaches.
Using the standard bosonization method, 
   Fabrizio, Gogolin, and Nersesyan recently argued that
   the ground state of the ionic Hubbard model exhibits three phases
   as $U$ increases:
   the band insulator, the SDI,
   and the Mott insulator.\cite{Fabrizio}
The order parameter of the SDI state is 
   nothing but that of the BCDW state, and we can regard the two
   states as essentially identical.
It was also argued that the quantum phase
   transition from the band insulator to the SDI state belongs
   to the Ising universality class whereas the other transition
   from the SDI state to the Mott insulator is of the
   Kosterlitz-Thouless type.
Recent numerical studies,
   \cite{Takada,Qin,Brune,YZZhang,Anusooya-Pati,Wilkens,Torio,Refolio,Manmana}
   however, have reported
   controversial results on the existence of the SDI phase.
Some claimed to find two quantum phase transitions while others found
   evidences of only one phase transition.
With this issue of the SDI phase in mind,
    in this section we investigate the phase diagram of
    the 1D extended Hubbard model with the staggered site potential
    and examine critical properties of the quantum phase transitions.

We take into account the staggered site potential and the correlation
   effects on equal footing by treating them as weak perturbations.
We use Eq.\ (\ref{eq:field_op}) to rewrite $H_\Delta$ in the continuum
   limit as $H_\Delta=\int dx \, \mathcal{H}_\Delta$, where
   \cite{Fabrizio,Tsuchiizu_JPSJ}
\begin{equation}
\mathcal{H}_\Delta =
-\frac{g_\Delta}{2(\pi a)^2} \, \sin \theta \, \cos \phi
\label{eq:Hw_bosonization}
\end{equation}
   with $g_\Delta=4\pi \Delta a$.
Note that the CDW order parameter $\mathcal{O}_\mathrm{CDW}$ is
  proportional to $\mathcal{H}_\Delta$, and $g_\Delta$ can be regarded as
  an external force coupled to $\mathcal{O}_\mathrm{CDW}$.
This has the consequence that $\mathcal{O}_\mathrm{CDW}$ acquires
  a nonvanishing expectation value for any finite $U$ and $V$,
  as long as $g_\Delta\neq0$.
In this section we will denote the insulating phase connected to the
  free-electron band insulator ($U=V=0$ and $\Delta\ne0$) by
  the BI phase, rather than the CDW phase.

The bosonized form of the Hamiltonian $H'$ can be thought of as
   a generalization of the so-called double sine-Gordon (DSG) model
   as $H'$ contains sine/cosine terms
   with different frequencies ($\sin \theta$ and $\cos 2\theta$,
   $\cos \phi$ and $\cos 2\phi$). 
The DSG theory itself has been investigated intensively
   \cite{Fabrizio,Bajnok,Delfino} and shown to have a critical point
   belonging to the Ising universality class 
   [$c=\frac{1}{2}$ conformal field theory (CFT)].
To obtain a qualitative understanding of
   the critical properties in our system,
   we first perform a semiclassical analysis in a similar way 
   to Sec.\ III B, before examining the global phase diagram of $H'$
   with use of the RG method.

\subsection{Semiclassical analysis}

In this section, we perform a semiclassical analysis
   to the Hamiltonian $\mathcal{H}'=\mathcal{H}+\mathcal{H}_\Delta$,
   where $\mathcal{H}$ and $\mathcal{H}_\Delta$ are given by 
   Eqs.\ (\ref{eq:Hamiltonian}) and (\ref{eq:Hw_bosonization}),
   respectively.
We neglect spatial variations of the field
   and focus on the locking potential:
\begin{eqnarray}
V_\Delta(\theta,\phi)
&=&
-g_c \cos2\theta + g_s \cos 2\phi
-g_{cs} \cos 2\theta \cos 2\phi
\nonumber \\ 
&& {}
-g_\Delta \, \sin\theta \, \cos \phi
.
\label{eq:VW}
\end{eqnarray}

\begin{figure}[t]
\includegraphics[width=7cm]{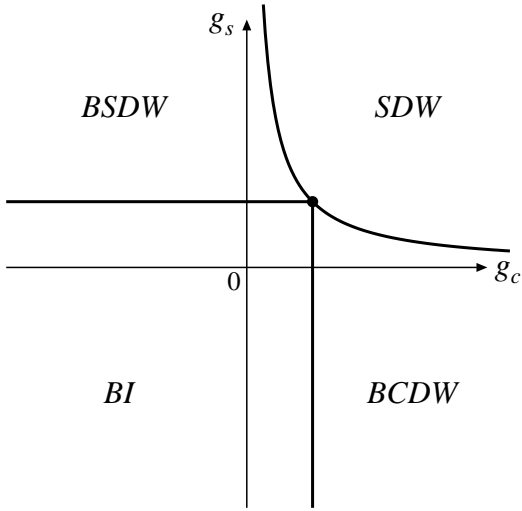}
\caption{
Phase diagram obtained by minimizing the potential energy 
   $V_\Delta^0(\theta,\phi)$ [Eq.\ (\ref{eq:VW0})].
The phase boundaries between the SDW state and 
   the BCDW state, and 
   between the SDW state and the BSDW state 
   are given by the curve $g_s=g_\Delta^2/(16 g_c)$ with $g_c>0$.
The phase boundaries between the BI state and 
   the BCDW state, and 
   between the BI state and the BSDW state 
   are given by the lines $g_c=\frac{1}{4}\,|g_\Delta|$
   with $g_s<\frac{1}{4}\,|g_\Delta|$ and $g_s=\frac{1}{4}\,|g_\Delta|$
   with $g_c<\frac{1}{4}\,|g_\Delta|$, respectively. 
All the phase transitions in this figure are continuous.
The tetracritical point is located at
   $(g_c,g_s)=
    (\frac{1}{4}\,|g_\Delta|,\frac{1}{4}\,|g_\Delta|)$.
}
\label{fig:classical-W}
\end{figure}

First, we examine the case $g_{cs}=0$, which corresponds to 
   the situation where the $g_{cs}$ term becomes irrelevant in the
   RG scheme.
The potential to be considered is
\begin{eqnarray}
V^0_\Delta(\theta,\phi)& \equiv &
V_\Delta(\theta,\phi)|_{g_{cs}=0} \nonumber\\
&=&
-g_c \, \cos2\theta + g_s \, \cos 2\phi
-g_\Delta \, \sin\theta \, \cos \phi
.
\nonumber\\&&
\label{eq:VW0}
\end{eqnarray}
Due to its double-frequency structure, possible locations
   of the phase locking are different from the ones we found
   in Sec.\ III B.
For example, when $g_c>0$ ($g_s>0$), 
   the two kinds of potentials proportional to
   $\sin \theta$ and $\cos 2\theta$ ($\cos \phi$ and $\cos 2\phi$)
   compete with each other. \cite{Delfino,Bajnok}
The locking of the phases $\theta$ and $\phi$
   are determined from the saddle-point equations: 
   $\cos\theta(4 g_c \sin \theta-g_\Delta \cos \phi)=0$ and
   $\sin\phi (- 4 g_s \cos \phi + g_\Delta \sin \theta)=0$.
In order to simplify the notations, let us introduce
\begin{equation}
\alpha_\theta^0
\equiv \left| \cos^{-1} \left(\frac{g_\Delta}{4g_c}\right) \right|
, \quad
\alpha_\phi^0
\equiv \left| \cos^{-1} \left(\frac{g_\Delta}{4g_s}\right) \right|
,
\end{equation}
   where $|g_\Delta/g_c|\le 4$, $|g_\Delta/g_s|\le 4$, and
   $0 \le \alpha_\theta^0,\alpha_\phi^0 \le \pi$ are assumed.
\begin{table}
\caption{Possible ordered ground states  and
  the position of (quasi-)locked phase fields
  determined from  Eq.\ (\protect{\ref{eq:VW0}}).
}
\label{table:phase-locking-W}
\begin{ruledtabular}
\begin{tabular}{lc}
Phase  &  $(\theta,\phi )$       \\
\hline
SDW    &  $(0,\pm \pi/2 ), (\pi,\pm \pi/2 )$  \\
BI (for $g_\Delta>0$)   &  
    $(+\pi/2,0 ), (-\pi/2,\pi )$  \\
BI (for $g_\Delta<0$)   &
    $(+\pi/2,\pi ),(-\pi/2,0 )$  \\
BCDW & 
    $\biglb(+(\pi/2)\pm \alpha_\theta^0,0 \bigrb), 
     \biglb(-(\pi/2)\pm \alpha_\theta^0,\pi \bigrb)$  \\
BSDW & 
   $(+\pi/2, 0\pm \alpha_\phi^0 ),
      \biglb(-\pi/2, \pm (\pi-\alpha_\phi^0) \bigrb)$
\end{tabular}
\end{ruledtabular}
\end{table}
\begin{figure}[b]
\includegraphics[width=7.cm]{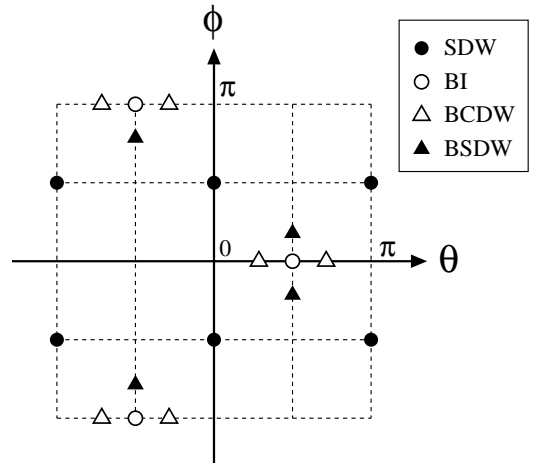}
\caption{
Positions of locked phase fields $\theta$ and $\phi$
   in the four states when $g_\Delta>0$. 
}
\label{fig:potmin-W}
\end{figure}%
The solutions of the saddle-point equations yield
   the following four states with distinct
   configurations of the locked phase fields $\theta$ and $\phi$
   (modulo $2\pi$):
(i) the SDW state with $\theta$ and $\phi$ locked at
   $(\theta,\phi)=(0,\pm \pi/2)$ or 
   $(\pi, \pm \pi/2)$;
(ii) the BI state with
   $(\theta,\phi) =(+\pi/2,0)$, $(-\pi/2,\pi)$
   if $g_\Delta>0$ and with
   $(\theta,\phi)=(+\pi/2,\pi)$, $(-\pi/2,0)$
   if $g_\Delta<0$;
(iii) the ``BCDW'' state where the BCDW order and the CDW order
   coexist and which is realized when
   $(\theta,\phi)
   =( \pi/2\pm \alpha_\theta^0,0)$ or
   $(-\pi/2\pm \alpha_\theta^0,\pi)$;
(iv) the ``BSDW'' state where the BSDW and the CDW order coexist
   and which is realized when
   $(\theta,\phi)
    =(\pi/2,0\pm \alpha_\phi^0)$ or
    $\biglb(-\pi/2, \pm( \pi - \alpha_\phi^0) \bigrb)$.
Table \ref{table:phase-locking-W} and Fig.\ \ref{fig:potmin-W}
  summarize the possible ordered ground states and corresponding
  positions of locked phase fields.
The potential energies in these states are given by
\begin{subequations}
\begin{eqnarray}
V^0_{\rm SDW}
&=& -g_c -g_s 
,
\\
V^0_{\rm BI}
&=&  + g_c + g_s - \left| g_\Delta \right|
,
\\
V^0_{\rm BCDW}
&=&
-g_c +g_s -\frac{g_\Delta^2}{8g_c} 
,
\label{eq:V0_BCDW+CDW}
\\
V^0_{\rm BSDW}
&=&
 + g_c -g_s 
 - \frac{g_\Delta^2}{8g_s}
.
\label{eq:V0_BSDW+CDW}
\end{eqnarray}
\end{subequations}
In deriving Eqs.\ (\ref{eq:V0_BCDW+CDW}) and (\ref{eq:V0_BSDW+CDW}), 
   we have assumed $|g_\Delta/g_c| \le 4$ and $|g_\Delta/g_s| \le 4$,
   respectively.
The CDW state is stabilized strongly by the $g_\Delta$ term
   whereas
   the BCDW state and the BSDW state are also stabilized by
   the second-order contribution of $g_\Delta$.
By comparing these energies, we arrive at the phase diagram
   shown in Fig.\  \ref{fig:classical-W}.
As we go across the boundary ($g_c=\frac{1}{4}\, g_\Delta$) 
   from the BI state to the BCDW state,
   we find that each potential minimum splits into 
   two minima, e.g., $( \theta, \phi )
    =( \pi/2,0 ) \to 
     \biglb( (\pi/2)\pm \alpha_\theta^0, 0 \bigrb)$,
   and that
   the potential for the $\theta$ phase field takes a
   double-well structure in the BCDW state.
Similarly, as we go from the BI state to the BSDW state,
   each potential minimum splits into 
   two minima, e.g., $( \theta, \phi )
    =(\pi/2,0) \to ( \pi/2, \pm \alpha_\phi^0)$,
   and now the potential for the $\phi$ phase field has a
   double-well structure in the BSDW state.
As long as $g_{cs}=0$, any quantum phase transition is continuous
   since a potential barrier between two potential minima
   corresponding to two different states vanishes at the transition.
The phase diagram (Fig.\ \ref{fig:classical-W}) indicates that
   a direct transition 
   from the SDW state to the BI state takes place only when 
   the parameters $g_c$ and $g_s$ are on the multicritical point
   $(g_c,g_s)
     =(\frac{1}{4}\, |g_\Delta|,\frac{1}{4}\,|g_\Delta|)$,
   where the potential takes the form
   $V_\Delta^0(\theta,\phi)=
    \frac{1}{2}\, |g_\Delta| \{ -1 
   + \left[\sin\theta - \mathrm{sgn}(g_\Delta) \cos\phi \right]^2 \}$ 
   and is minimized
   at $\phi=\pm [(\pi/2) -\theta]$ 
   and $\phi=\pm (\frac{3}{2}\pi +\theta)$ if $g_\Delta>0$, or
   at $\phi=\pm [(\pi/2) +\theta]$ 
   and $\phi=\pm (\frac{3}{2}\pi -\theta)$ if $g_\Delta<0$. 

Let us take a closer look at low-energy excitations
   in the BI state and the BCDW state.
The massive sine-Gordon model has
   topological excitations, solitons, and antisolitons.
They are characterized by the
   topological charges $Q$ and $S_z$
    for the charge and the spin sectors,
\begin{equation}
Q = \frac{1}{\pi} \int dx \, \partial_x \theta 
, \quad
S_z = \frac{1}{2\pi} \int dx \, \partial_x \phi
.
\label{eq:topological_charge}
\end{equation}
In the noninteracting case ($U=V=0$) with a finite $\Delta$,
   the lowest-energy excitation is a soliton 
   of $\theta$ and $\phi$ connecting two neighboring minima of
   the $-g_\Delta\sin\theta \, \cos \phi$, e.g., 
   $(\theta,\phi)|_{x\to-\infty}=(-\pi/2,\pi)$
   and $(\theta,\phi)|_{x\to\infty}=(\pi/2,0)$.
Such an excitation carries the charge $Q=\pm 1$ and 
   the spin $S_z=\pm \frac{1}{2}$, 
   which is nothing but a single-electron excitation
   in the band insulator. 
It has been pointed out \cite{MJRice,Nagaosa,Fabrizio} that
   in the SDI phase (i.e., in the BCDW phase),
   the topological charge $Q$ of the lowest-energy excitation
   becomes fractional, $Q=\pm 2\alpha_\theta^0/\pi$, 
   reflecting the local double-well structure of the potential
   near the potential minima, e.g., at
   $(\theta,\phi)=(\pi/2\pm\alpha_\theta^0,0)$.
This is a unique feature of the BCDW phase and is contrasted 
   from the integer charge $Q=\pm 1$ of the lowest-energy
   excitation in the pure BCDW phase where the phase fields are
   locked at $(\theta,\phi)=(0,0)$. 
Accordingly, the phase transition between the BCDW state
   and the BI state belongs to a different universality class from 
   the one between the pure BCDW state and the CDW state
   discussed in Sec.\ \ref{sec:SDW-CDW_trans}.
In the former case, a small potential barrier in a double-well
   potential in the BCDW state vanishes at the critical point
   and the effective theory for the low-energy excitations
   is the ``$\varphi^4$'' theory known to describe
   the Ising phase transition, rather than the 
   Gaussian theory that governs the transition
   between the BCDW and CDW phases.

One might expect that a similar semiclassical analysis can be
   applied to the spin field $\phi$.
Within the semiclassical approach the topological charge $S_z$
   in the BSDW phase of Fig.\ \ref{fig:classical-W} takes a
   fractional value, $\pm \alpha_\phi^0/(2\pi)$. 
However, 
   since the Hamiltonian has the global SU(2) spin-rotation symmetry,
   the SDW state and the BSDW state cannot have a true long-range order.
This implies that the phase field $\phi$ cannot be localized
   except in spin-gap phases
   where $\phi$ is locked at $\langle\phi\rangle=0$ mod $\pi$. 
The global SU(2) symmetry thus prohibits the Ising criticality
   in the spin sector.
In fact, the BSDW phase in Fig.\ \ref{fig:classical-W} 
   turns out to be just the BI phase.

\begin{figure}[t]
\includegraphics[width=7.2cm]{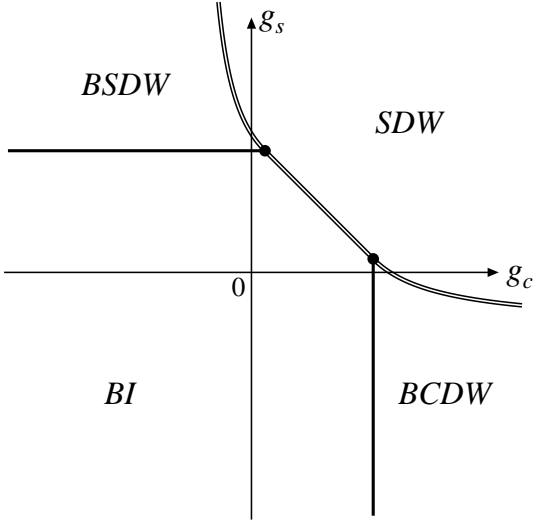}
\caption{
Phase diagram obtained by minimizing the potential energy 
   $V_\Delta(\theta,\phi)$ [Eq.\ (\ref{eq:VW})]
   for $g_{cs}<0$.
The phase boundaries are given by 
   $g_s=-g_c+\frac{1}{2}\,|g_\Delta|$ 
   between the SDW and the BI states,
   $g_c=|g_{cs}|+\frac{1}{4}\,|g_\Delta|$
   between the BI and the BCDW states,
   $g_s=|g_{cs}|+\frac{1}{4}\,|g_\Delta|$
   between the BI and the BSDW states,
   $g_s=-|g_{cs}|+g_\Delta^2/[16(g_c-|g_{cs}|)]$
   between the SDW and the BCDW states, and 
   $g_c=-|g_{cs}|+g_\Delta^2/[16(g_s-|g_{cs}|)]$
   between the SDW and the BSDW states.
Multicritical points are located at 
   $(g_c,g_s)
   =(+|g_{cs}|+\frac{1}{4}\, |g_\Delta|,
           -|g_{cs}|+\frac{1}{4}\, |g_\Delta|)$ and 
   $(-|g_{cs}|+\frac{1}{4}\, |g_\Delta|,
           +|g_{cs}|+\frac{1}{4}\, |g_\Delta|)$.
The single lines denote second-order transitions, while 
   the double lines denote first-order transitions. 
}
\label{fig:classical-W2}
\end{figure}

Let us now consider the situation in which $g_{cs}\neq 0$.
In this case, the phase fields $\theta$ and $\phi$ are locked
   in a similar way to the case $g_{cs}=0$, 
   but $\alpha_\theta^0$ and $\alpha_\phi^0$ are 
   modified into $\alpha_\theta^0\to \alpha_\theta$ and
   $\alpha_\phi^0\to \alpha_\phi$, where
\begin{subequations}
\begin{eqnarray}
\alpha_\theta
&\equiv& 
\left| \cos^{-1} \left[\frac{g_\Delta}{4(g_c-|g_{cs}|)}\right] \right|
, \,\,\,
\\
\alpha_\phi
&\equiv& 
\left| \cos^{-1} \left[\frac{g_\Delta}{4(g_s-|g_{cs}|)}\right] \right|.
\end{eqnarray}
\end{subequations}
Here we have assumed
   $|g_\Delta/(g_c-|g_{cs}|)|\le 4$ and
   $|g_\Delta/(g_s-|g_{cs}|)|\le 4$.
The potential energies in the four states become
\begin{subequations}
\begin{eqnarray}
V_{\rm SDW} &=& -g_c -g_s - |g_{cs}|  ,  \\
V_{\rm BI} &=& +g_c + g_s - |g_{cs}|  -|g_\Delta|  , \\
V_{\rm BCDW} 
&=&
-g_c +g_s + |g_{cs}| 
-\frac{g_\Delta^2}{8(g_c-|g_{cs}|)}
, \qquad 
\label{eq:V_BCDW+CDW}
\\
V_{\rm BSDW}
&=&
 + g_c -g_s + |g_{cs}|  
-\frac{g_\Delta^2}{8(g_s-|g_{cs}|)}
. \quad 
\label{eq:V_BSDW+CDW}
\end{eqnarray}
\end{subequations}
By comparing these energies we obtain
   the phase diagram (Fig.\ \ref{fig:classical-W2}).
In the limit $g_\Delta\to 0$ this phase diagram reduces to 
   Fig.\ \ref{fig:classic}.
One can easily find that the $g_{cs}$ term favors the 
   SDW state and the BI state over the
   BCDW state and the BSDW state. 
The direct SDW-BI transition line acquires a finite length
   in the phase diagram, like in Fig.\ \ref{fig:classic}.
The analysis of critical properties of each quantum phase transition
   is more complicated than that in Sec.\ III
   due to the presence of two kinds of charge-spin coupled terms, 
   the $g_\Delta$ and $g_{cs}$ terms.
Along the phase boundary between the SDW state and the BI state,
   the potential energy is minimized at discrete points,
   $(\theta,\phi)=(-\pi/2, \pi)$, $(0,\pm \pi/2)$,
   $(\pi/2,0)$, $(\pi,\pm \pi/2)$ for $g_\Delta>0$, or
   at $(\theta,\phi)=(-\pi/2, 0)$, $(0,\pm \pi/2)$,
   $(\pi/2,\pi)$, $(\pi,\pm \pi/2)$ for $g_\Delta<0$.
These points correspond either to
   the SDW state or to the BI state 
   (see Table \ref{table:phase-locking-W}).
Since any path connecting these potential minima has to go over
   a potential barrier, the direct SDW-BI transition is first order.
In addition, both the transition between the SDW state and
   the BCDW state and that between the SDW state and the 
   BSDW state become first order when $g_{cs}\ne0$.
On the phase boundary between the SDW state and the BCDW state,
   the potential has isolated minima
   at $(\theta,\phi) =(0,\pm \pi/2)$,
    $(\pi,\pm \pi/2)$,
    $(-\pi/2\pm \alpha_\theta, \pi)$, and 
    $(+\pi/2\pm \alpha_\theta, 0)$.
The pinning of the phase fields at these minima corresponds either to 
   the SDW state or to the BCDW state 
   (see Fig.\ \ref{fig:potmin-W}).
On the multicritical points at
   $(g_c,g_s)
   =(+|g_{cs}|+\frac{1}{4}\, |g_\Delta|,
           -|g_{cs}|+\frac{1}{4}\, |g_\Delta|)$ and 
   $(-|g_{cs}|+\frac{1}{4}\, |g_\Delta|,
           +|g_{cs}|+\frac{1}{4}\, |g_\Delta|)$, 
   the potential takes the form 
\begin{subequations}
\begin{eqnarray}
V_\Delta^{c1}(\theta,\phi)&=&
  -  |g_{cs}| (\cos 2\theta + \cos 2\phi - \cos 2\theta \, \cos 2\phi)
\nonumber \\ && {}
  + \frac{1}{2}\, |g_\Delta| 
    \left\{-1 + 
    \left[\sin\theta - \mathrm{sgn}(g_\Delta)\cos\phi \right]^2 \right\}
,
\nonumber \\
\label{eq:Vc1}
\\
V_\Delta^{c2}(\theta,\phi)&=&
  +  |g_{cs}| (\cos 2\theta + \cos 2\phi + \cos 2\theta \, \cos 2\phi)
\nonumber \\ && {}
  + \frac{1}{2}\, |g_\Delta| 
    \left\{-1 + 
    \left[\sin\theta - \mathrm{sgn}(g_\Delta)\cos\phi \right]^2 \right\}
,
\nonumber \\
\label{eq:Vc2}
\end{eqnarray}
\label{eq:Vc}%
\end{subequations}
   respectively.
The potential minima of $V_\Delta^{c1}(\theta,\phi)$ and
   $V_\Delta^{c2}(\theta,\phi)$
   are located at 
   $(\theta,\phi)=(-\pi/2, \pi)$, 
   $(0,\pm \pi/2)$,
   $(\pi/2,0)$, and
   $(\pi,\pm \pi/2)$ for $g_\Delta>0$ and
   at $(\theta,\phi)=(-\pi/2, 0)$, 
   $(0,\pm \pi/2)$,
   $(\pi/2,\pi)$, and
   $(\pi,\pm \pi/2)$ for $g_\Delta<0$.

Finally, we note that even in the SDW state (the Mott insulator)
the CDW order parameter has a nonvanishing expectation value.
This is because the alternating site potential $\mathcal{H}_\Delta$
has the same form as the CDW order parameter
$\mathcal{O}_\mathrm{CDW}\propto\sin\theta\cos\phi$.
Even though the semiclassical analysis indicates that the phase
fields are pinned, say, at $(\theta,\phi)=(0,\pm\pi/2)$, quantum
fluctuations of the fields around the pinning position lead to
a nonvanishing $\langle\mathcal{O}_\mathrm{CDW}\rangle$.
This can be easily seen in the limit of small $\Delta$, where
\begin{eqnarray}
\langle\mathcal{O}_\mathrm{CDW}\rangle&\propto&
\mathrm{Tr}\left[
  \exp\left[-\int dx(\mathcal{H}+\mathcal{H}_\Delta)\right]
  \sin\theta\cos\phi
           \right]
\nonumber\\
&\propto&
g_\Delta\mathrm{Tr}\left[\exp\left(-\int dx\mathcal{H}\right)
\sin^2\theta\cos^2\phi\right]
\ne0.
\nonumber \\
\end{eqnarray}

\subsection{Renormalization-group analysis}

We perform RG analysis to take into account quantum fluctuations
   that are ignored in the semiclassical analysis.
As in Sec.\ III, we obtain the RG
   equations using the OPE method
   (see Appendix \ref{sec:rg}): 
\begin{eqnarray}
\frac{d}{dl} G_\Delta 
  \!\!&=&\!\! {} + G_\Delta
    + \frac{1}{2} \, G_\Delta \, G_\rho 
    - G_\Delta \, G_c
\nonumber \\ && {}
    - \frac{3}{2} \, G_\Delta \, G_s
    - \frac{3}{4} \, G_\Delta \, G_{cs}
    - \frac{3}{8} \, G_\Delta \, G_{\rho s}
,\quad\quad
\label{eq:RGf-b}
\\
\frac{d}{dl} G_\rho 
  \!\!&=&\!\! {} + \frac{1}{4} \, G_\Delta^2
    + 2 \, G_c^2 + G_{cs}^2 +  G_s \, G_{\rho s}
,
\\
\frac{d}{dl} G_c 
  \!\!&=&\!\! {}
  - \frac{1}{4} \, G_\Delta^2 
    + 2 \, G_\rho\, G_c
  - G_s \, G_{cs} - G_{cs} \, G_{\rho s}
,  \quad\quad
\\
\frac{d}{dl} G_s 
  \!\!&=&\!\! {} - \frac{1}{4} \, G_\Delta^2 
    - 2 \, G_s^2 - G_c \, G_{cs} - G_{cs}^2
,
\label{eq:RGf-gs}
\\
\frac{d}{dl} G_{cs} 
  \!\!&=&\!\! {} - \frac{1}{4}  \,G_\Delta^2 
    - 2 \, G_{cs} + 2 \, G_\rho \, G_{cs} - 4 \, G_s \, G_{cs}
\nonumber \\ && {}
    - 2 \, G_c\, G_{s}
    - 2 \, G_c\, G_{\rho s}
    - 4 \, G_{cs} \, G_{\rho s}
,
\\
\frac{d}{dl} G_{\rho s} 
  \!\!&=&\!\! {} - \frac{1}{4}  \, G_\Delta^2 
    - 2 \, G_{\rho s} + 2 \, G_\rho \, G_s
\nonumber \\ && {}
    - 4 \, G_c \, G_{cs} - 4 \, G_{cs}^2
    - 4 \, G_s \, G_{\rho s}
.
\label{eq:RGf-e}
\end{eqnarray}
The initial value of $G_\Delta(l)$ is given by 
   $G_\Delta(0)=\Delta/t$, while those of the other coupling
   constants are given by $G_\nu(0) =g_\nu /(4\pi ta)$.
Since the RG equations are invariant under the sign change of
   $G_\Delta$ ($G_\Delta \to -G_\Delta$), we can assume 
   $G_\Delta(0)\ge 0$ without losing generality
   in the following arguments.

We determine the ground-state phase diagram
   in a similar way as in Sec.\ \ref{sec:phase_diagram}.
That is, we integrate the scaling equations 
   (\ref{eq:RGf-b})--(\ref{eq:RGf-e}) numerically
   and find which one of the couplings
   [$G_\Delta(l)$, $G_c(l)$, $G_s(l)$, and $G_{cs}(l)$]
   becomes most relevant.
By doing so, we have encountered the following four cases.

(i) The case where $G_c(l)$ grows fastest and becomes $1$
   at $l=l_{\rho+}$.
Below this energy scale (i.e., $l\ge l_{\rho+}$),
   the charge fluctuations are suppressed and 
   the phase field $\theta$ is locked at $\theta = 0$ or $\pi$.
For the discussion of the ground-state properties
   we may first neglect the $g_\Delta$ term 
   since $\langle \sin \theta \rangle \cos \phi=0$.
The Hamiltonian density $\mathcal{H}'$ then reduces to
\begin{eqnarray}
\mathcal{H}^{\mathrm{eff}}_{\sigma +} &=&
\frac{v_F}{2\pi} \sum_p (\partial_x \phi_p)^2
- \frac{v_F}{\pi} \, G_s^* 
   \left(\partial_x \phi_+\right) \left(\partial_x \phi_-\right) 
\nonumber \\ && {}
+\frac{v_F}{\pi a^2}  \, G_s^* \cos 2\phi
,
\end{eqnarray}
   where $G_s^* = G_s(l_{\rho+}) - G_{cs}(l_{\rho+})$.
We immediately see that, if $G_s^*>0$, the spin excitations are  
   gapless and the ground state is the SDW state.
On the other hand, if $G_s^*<0$, then 
   the operators proportional to $G_s^*$ are relevant
   [$G_s^*(l)\to -\infty$ under scaling]
   and the phase fields are locked as 
   $(\theta,\phi )=(0,0),
   (0,\pi),(\pi,0),(\pi,\pi)$,   
   which corresponds to the BCDW state with
   $\alpha_\theta \to \pi/2$
   (i.e., $g_\Delta \to 0$),
   see Table \ref{table:phase-locking-W}.
This would become the BCDW state with $\alpha_\theta<\pi/2$
   in a more realistic treatment where the $g_\Delta$ term is not
   simply ignored.

(ii) The case where $|G_c(l)|$ grows most rapidly and $G_c(l)\to -1$
   at $l=l_{\rho-}$.
The phase field $\theta$ is then locked at $\theta=\pm \pi/2$
   for $l>l_{\rho-}$.
Below this energy scale one can replace the $\sin\theta$ potential 
  by its averaged value, i.e., $\sin \theta \to \langle\sin \theta
  \rangle = \pm 1$.
The effective Hamiltonian at $l=l_{\rho-}$ is given by
\begin{eqnarray}
\mathcal{H}^{\mathrm{eff}}_{\sigma -} &=& 
\frac{v_F}{2\pi} \sum_p (\partial_x \phi_p)^2
- \frac{v_F}{\pi} \, G_s^* 
   \left(\partial_x \phi_+\right) \left(\partial_x \phi_-\right) 
\nonumber \\ && {}
\mp \frac{v_F}{\pi a^2} \, G_\Delta^* \, \cos \phi
+ \frac{v_F}{\pi a^2} \, G_s^* \, \cos 2\phi
,
\label{eq:Heff_sigma-}
\end{eqnarray}
   where $G_\Delta^*=G_\Delta(l_{\rho-})$ and
   $G_s^*=G_s(l_{\rho -})+G_{cs}(l_{\rho-})$, 
   and the sign $-/+$ 
   of the $G_\Delta$ term corresponds to the position of
   the phase locking $\theta=+ (\pi/2)/- (\pi/2)$.
When $G_s^*>0$, the two $G_s^*$ terms are 
   marginally irrelevant,
   and the only relevant operator is $\mp \cos\phi$. 
Then the phase field $\phi$ is locked at $\phi=0$ or $\pi$,
   depending on the position of
   the charge phase locking $\theta=+ (\pi/2)$ or 
   $- (\pi/2)$.
On the other hand, when $G_s^*<0$,
   both $G_\Delta^*$ and $G_s^*$ terms become relevant. 
However, these terms do not compete with each other.
The only effect of the $G_\Delta^*$ term is to lift the degeneracy
   between the neighboring minima of $-\cos2\phi$,
   and hence the position of the phase locking is the same as
   in the case $G_s^*>0$. 
Therefore, regardless of the sign of $G_s^*$,
   the resultant phase is found to be the BI state
   with the phase locking at
   $(\theta,\phi)=(\pi/2,0)$ or $(-\pi/2,\pi)$.

(iii)
The case where either $|G_{cs}(l)|$ or $|G_\Delta(l)|$ is most relevant.
Then both charge and spin fluctuations are suppressed, and 
   the classical treatment is sufficient at lower energy scale.
In this case, we find to which phase the ground state belongs
   by substituting the parameters $G_c(l)$ and $G_s(l)$ into $g_c$
   and $g_s$ in Fig.\ \ref{fig:classical-W2}.

(iv) 
The case where $G_{s}(l)$ is most relevant and becomes $-1$
   at $l=l_\sigma$. 
Below this energy scale the spin fluctuations are suppressed and 
   the phase field $\phi$ is locked as $\phi \to 0$ or $\pi$ 
   for $l>l_\sigma$.
The effective Hamiltonian of the remaining charge sector is
\begin{eqnarray}
\mathcal{H}^{\mathrm{eff}}_{\rho}
&=&
\frac{v_F}{2\pi} \sum_p (\partial_x \theta_p)^2
+ \frac{v_F}{\pi} \, G_\rho^*
  \left(\partial_x \theta_+ \right)
  \left(\partial_x \theta_- \right)
\nonumber \\ && {}
\mp \frac{v_F}{\pi a^2} \, G_\Delta^* \, \sin \theta
-\frac{v_F}{\pi a^2} \, G_c^* \, \cos 2\theta, 
\end{eqnarray}
   where $G_\rho^*=G_\rho(l_\sigma)-G_{\rho s}(l_\sigma)$,
   $G_\Delta^*=G_\Delta(l_\sigma)$, and
   $G_c^*=G_c(l_\sigma)+G_{cs}(l_\sigma)$. 
The sign $-/+$ of the $G_\Delta^*$ term 
   corresponds to the position of the phase locking $\phi=0/\pi$. 
In this Hamiltonian, both of the nonlinear terms, $\sin \theta$
   and $\cos 2\theta$, are relevant operators.
If $G_c^*<0$, then the situation is the same as the case (ii):
   the $G_\Delta^*$ and $G_c^*$ terms do not compete with each other and
   the possible phase locking pattern is
   $\theta=+\pi/2$ $(-\pi/2)$
   for $\phi=0$ $(\pi)$, where the ground state is the BI state.
If $G_c^*>0$, these two terms compete with each other, since
   the $-(+) \sin \theta$ potential tends to lock the phase field 
   $\theta$ at $\theta=+\pi/2$ $(-\pi/2)$,
   while the $\cos 2\theta$ potential tends to lock it at $\theta=0$ or 
   $\pi$.
In this case, possible ground states are
   the BI state and the BCDW state, and 
   the quantum phase transition between them is
   of the Ising transition type with the central charge $c=1/2$,
   as discussed in the preceding section. 
However, it is hard to estimate quantitatively the critical value of
   the coupling constants at 
   the quantum phase transition.
One way to estimate it is to find a critical point separating the
   basins of attraction to the two strong-coupling fixed points,
   $(G_\Delta^*,G_c^*)\to (+ \infty,-\infty)$ and
   $(0,+\infty)$, in the perturbative RG analysis.
   \cite{Tsuchiizu_JPSJ,Tsuchiizu_JPSJ2}
However, with this method where the cosine and sine terms are treated
   perturbatively, we cannot see the correct picture of the DSG
   theory with the double-well potential structure which leads to the
   Ising transition.
Instead, here we estimate the critical value for the Ising transition
   from the semiclassical arguments:
   The critical value is determined from the condition
   $G_c^*/G_\Delta^*=1/4$.

\begin{figure}[t]
\includegraphics[width=7cm]{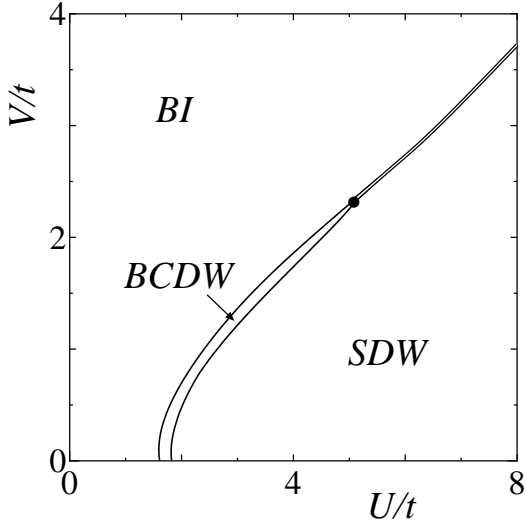}
\caption{
Phase diagram of the half-filled extended Hubbard model
 at $\Delta/t=0.1$.
The double line denotes the first-order transition,
  while the single lines denote the second-order transitions.
}
\label{fig:UVW}
\end{figure}

\begin{figure}[b]
\includegraphics[width=7cm]{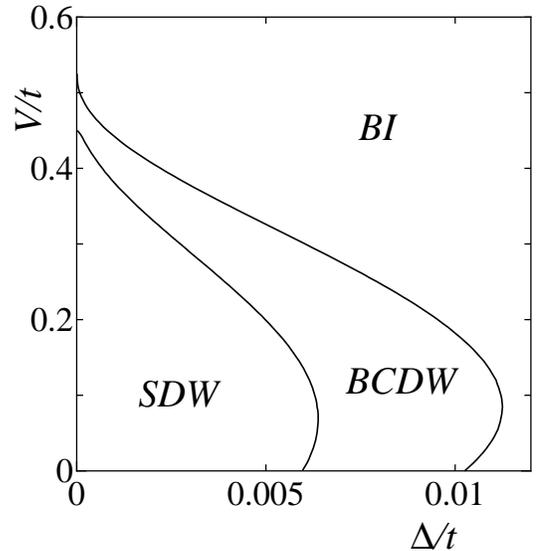}
\caption{
Phase diagram of the half-filled extended ionic Hubbard model 
  on the plane of $\Delta/t$ and $V/t$, where $U/t=1$.
}
\label{fig:UVW-wv}
\end{figure}

We have used the above scheme to obtain the phase diagram
   shown in Fig.\ \ref{fig:UVW}, for which $\Delta/t=0.1$.
The phase diagram at large $U$ and $V$ is similar to
   Fig.\ \ref{fig:phase2}, whereas
   a qualitative charge in the phase diagram is found in
   the region $U,V \lesssim t$.
In agreement with Fabrizio, Gogolin, and Nersesyan,\cite{Fabrizio}
   we obtain two critical points ($U_{c1}<U_{c2}$)
   separating three phases on the $U$ axis:
   the BI state, the BCDW state (= the SDI state\cite{Fabrizio}),
   and the SDW state.
From comparison of Figs.\ \ref{fig:phase2} and \ref{fig:UVW}, 
   we see that the BCDW state in Fig.\ \ref{fig:phase2} has evolved
   continuously into the BCDW state when the alternating site potential
   $\Delta$ is switched on.
The phase diagram in the $\Delta$-$V$ plane is shown in Fig.\
   \ref{fig:UVW-wv}, where $U/t=1$.
Both $\Delta$ and $V$ promote the BI state, while 
   the SDW ground state is obtained for small $\Delta (\ll U)$
   and $V(\ll U)$.
We find that the region of the BCDW state obtained in the EHM
   at $\Delta=0$
   is connected to the region of the BCDW state in the Hubbard model
   with alternating site potential at $V=0$.

Let us discuss in more detail the critical regime
   in the limit of small $U$, $V$, and $\Delta$.
In this region we can safely neglect the irrelevant terms and set
   $G_{cs}(l)=G_{\rho s}(l)=0$ in the RG equations 
   (\ref{eq:RGf-b})--(\ref{eq:RGf-e}).
First we consider the case $V=0$.
Integrating out the RG equations 
   (\ref{eq:RGf-b})--(\ref{eq:RGf-gs}) analytically 
and following the criterion discussed above,
   we obtain asymptotic expansion of 
   the critical values for small $\Delta/t$:
\begin{eqnarray}
U_{c1}^{0} \!\! &=& \!\!
\frac{2\pi t}{\ln (t/\Delta)}
\left[1 - \frac{C}{\ln(t/\Delta)}+\cdots\right],
\\
U_{c2}^{0} \!\! &=& \!\!
\frac{2\pi t}{\ln (t/\Delta)}
\left[1 + C'\frac{\ln\ln(t/\Delta)}{\ln(t/\Delta)} 
                     + O\left(\frac{1}{\ln(t/\Delta)}\right) \right],
\nonumber \\
\end{eqnarray}
where $C$ and $C'$ are positive constants of order unity. 
The $\Delta$ dependence of $U_{c1}^{0}$ is different from 
   the result in Refs.\ \onlinecite{Fabrizio} since
   the lowest correction to $2\pi t/\ln(t/\Delta)$ 
   is not $O\biglb(\ln[\ln(t/\Delta)]/\ln(t/\Delta)\bigrb)$, but
   $O\biglb(1/\ln(t/\Delta)\bigrb)$.
Our results suggest that the ratio of $U_{c2}^{0}$ to
   $U_{c1}^{0}$ becomes
   $U_{c2}^{0}/U_{c1}^{0}=1+C' \ln[\ln(t/\Delta)]/\ln(t/\Delta)$.
At present we do not know where this difference comes from.
We extend this analysis to the case with finite $V(\ll U)$
   and examine the $V$ dependence of $U_{c1}$ and $U_{c2}$.
We note that $G_\rho(l) \neq G_c(l)$ in this case since
   the SU(2) symmetry of the charge sector is broken.
We integrate the RG equations analytically for small $V\neq 0$
   and obtain the corrections to order $V$,
\begin{subequations}
\label{U_c}
\begin{eqnarray}
U_{c1}
\!\! &=& \!\!
U_{c1}^0
-V\left[\frac{2}{3}+O\left(\frac{1}{\ln(t/\Delta)}\right)\right]
, 
\\
U_{c2} 
\!\! &=& \!\! 
U_{c2}^0
-V\left[\frac{2}{3}
        +O\left(\frac{\ln\ln(t/\Delta)}{\ln(t/\Delta)}\right)\right]
,
\end{eqnarray}
\end{subequations}
implying that the BCDW state survives upon inclusion of
   the $V(\ll U)$ term. 
We note that 
   $U_{c1}$ and $U_{c2}$ have a similar linear dependence on $V$.
From Eqs.\ (\ref{U_c}) and 
   Figs.\ \ref{fig:UVW} and \ref{fig:UVW-wv}, 
   we conclude that the phase diagram exhibits reentrant
   behavior as $V$ increases from zero with $\Delta$ and $U$
   being fixed at values near a quantum critical point.

\begin{figure}[t]
\includegraphics[width=8.5cm]{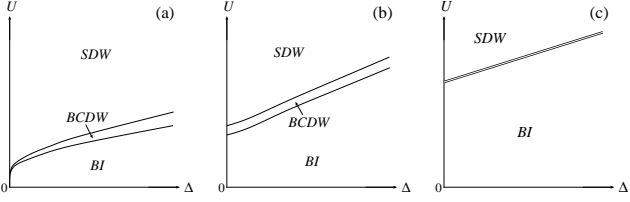}
\caption{
Schematic phase diagram of the half-filled extended Hubbard model
 at (a) $V=0$, (b) $V\ll t$, and (c) $V\gg t$.
The single lines represent second-order transitions, and the
 double line in (c) represents a first-order transition.
}
\label{fig:schematicWU}
\end{figure}

Since the Hamiltonian $H'$ has three free parameters ($U/t$, $V/t$,
and $\Delta/t$) at half filling, the ground-state phase diagram becomes
a three-dimensional (3D) diagram.
Instead of drawing such a 3D plot, here we show two-dimensional
tomographic phase diagrams.
Figure \ref{fig:schematicWU} shows schematic phase diagrams in the
$\Delta$-$U$ plane for three typical cases $V/t=0$, $V/t\ll1$, and
$V/t\gg1$.
We see that the nearest-neighbor repulsion enhances the BI phase
and destroys the BCDW phase at large $V$, where the direct transition
between the BI and SDW phases is first order.
The recent numerical study of the ionic Hubbard model\cite{Torio}
reports a similar phase diagram as Fig.\ \ref{fig:schematicWU}(a).
The first-order transition line in Fig.\ \ref{fig:schematicWU}(c)
asymptotically approaches the line $U=2\Delta+2V$.

\begin{figure}[t]
\includegraphics[width=8.5cm]{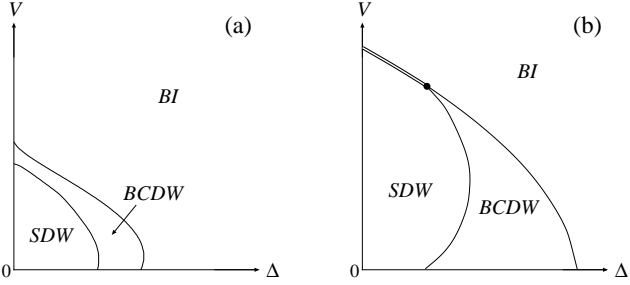}
\caption{
Schematic phase diagram of the half-filled extended Hubbard model
 at (a) $U\ll t$ and (b) $U\gg t$.
The single lines represent second-order transitions, and the
 double line represents a first-order transition.
}
\label{fig:schematicWV}
\end{figure}

Figure \ref{fig:schematicWV} shows schematic phase diagrams in
the $\Delta$-$V$ plane for $U/t\ll1$ and $U/t\gg1$.
At large $U$ and $V$ there appears a direct first-order transition
between the BI and SDW phases in Fig.\ \ref{fig:schematicWV}(b).
This first-order transition is in agreement with 
the results obtained from the  strong-coupling analysis \cite{Nagaosa} 
and numerical calculations. \cite{Girlando,Yonemitsu}

\subsection{Discussions on previous numerical results}

As mentioned in Introduction, many groups have already reported
  on numerical studies of the ground-state phase diagram of the ionic
  Hubbard model.
Various numerical techniques were used in these studies, including
  the density-matrix renormalization-group (DMRG) method,
  \cite{Takada,Qin,Brune,YZZhang,Manmana}
  the quantum Monte Carlo method, \cite{Wilkens,Refolio}
  a finite-size cluster method, \cite{Anusooya-Pati} and 
  a level crossing analysis. \cite{Torio}
The main issue here is whether or not the SDI phase (BCDW phase) exists,
  and so far these numerical studies do not seem to have reached
  complete agreement yet.
Although most of recent studies report that the SDI phase
  appears near the boundary between the SDW phase and the BI phase,
  \cite{Qin,Brune,YZZhang,Wilkens,Torio,Refolio,Manmana}
  there are still some conflicting claims in the literature.
A less controversial issue\cite{noteUc2} is the determination of the second
  critical value $U_{c2}$ at which a spin gap closes
  and which can be estimated
  by computing the spin gap directly \cite{Takada,Qin}
   or by examining the BCDW order parameter.\cite{YZZhang,Manmana}
The determination of the critical point $U_{c1}$ and the critical
  behaviors around it are more controversial issues.
One way to estimate the critical value $U_{c1}$ is to use
   the complex parameter introduced by Resta and Sorrela.\cite{Resta}
Its diverging behavior at $U=U_{c1}$ indeed allows one to determine
   the critical point.\cite{Takada,Wilkens}
Another way to determine the critical point is to find a gap closing
  point in excitation spectra.
Since the charge sector is responsible for the quantum phase transition
   at $U=U_{c1}$, one might try to look at a charge gap directly.
However, numerical studies have found that a naive charge gap does not
   vanish at the critical point and is always finite.
Recent studies have shown\cite{Qin,Brune,Manmana} that the excitation gap
   that vanishes at $U=U_{c1}$ is the gap to the first excited state
   that has the same charge and spin quantum numbers as the ground state.
Let us discuss this point in more detail below.

In numerical studies,\cite{Takada,Qin,Brune}
   the ``charge gap'' $\Delta_c$ was \textit{defined} as
    $\Delta_c = E_0(L/2+1,L/2)+E_0(L/2-1,L/2)-2E_0(L/2,L/2)$,
   where $E_0(N_\uparrow,N_\downarrow)$ is the lowest energy 
   of a finite-size system with an even number of sites $L$ that has
   $N_\uparrow$ up-spin and $N_\downarrow$ down-spin electrons.
This quantity $\Delta_c$ measures the energy of the excitation with the 
   topological charge $Q=\pm 1$ and $S_z=\pm1/2$
   [Eq.\ (\ref{eq:topological_charge})], and is rather a single-electron
   excitation gap.
According to the bosonization theory (Sec.\ IV A),
   the charge transition at $U=U_{c1}$
   is described by the ``$\varphi^4$'' theory and is
   in the Ising universality class.
The transition occurs when 
   two degenerate local minima of the effective potential for the
   charge fields merge into a single local minimum.
As one approaches the transition point from the Ising ordered phase
   (that is, the SDI phase), the topological charge 
   $Q=\pm 2\alpha_\theta/\pi$ of a lowest-energy excitation is 
   decreasing to zero, while excitations with $Q=1$ remain massive.
Therefore the charge gap $\Delta_c$ does not vanish
   at this Ising critical point, and
   this quantum phase transition cannot be detected with $\Delta_c$.
Qin \textit{et al}.\ and Manmana \textit{et al}.\ also used
   $\Delta_e =  E_1\left(L/2,L/2\right) -E_0\left(L/2,L/2\right)$
   in their numerical analysis,
   where $E_1(N_\uparrow,N_\downarrow)$ is the energy of the first 
   excited state.\cite{Qin,Manmana}
The quantity $\Delta_e$ measures excited states with 
   the same number of electrons,
   whose total topological charge $Q=0$ in the sine-Gordon scheme.
In the Ising ordered phase,
   the first excited state with the topological charge $Q=0$
   would be a bound state (or breather) of a soliton
   with the topological charge $+2\alpha_\theta/\pi$ and an antisoliton
   with the charge $-2\alpha_\theta/\pi$, whose energy vanishes at
   the critical point.
On the other hand, in the Ising disordered phase near the critical point,
   the potential is almost flat and has very small curvature.
The low-energy excitations would then be small oscillations around
   potential minima (rather than soliton/antisoliton) whose energy
   approaches zero as $U\to U_{c1}-0$.
Thus the exciton gap $\Delta_e$ is a right measure to detect
   the quantum phase transition at $U=U_{c1}$.

\section{Effect of bond dimerization}\label{sec:dimer}

In this section, we consider the 1D EHM with staggered bond 
   dimerization,\cite{Su,Ortiz} i.e.,
   the Peierls modulation of the hopping matrix element.
The total Hamiltonian $H''$ 
   is given by $H''=H+H_\delta$, where $H$ is defined in
   Eq.\ (\ref{eq:H1D}) and
\begin{equation}
H_{\delta} = \delta \sum_{j,\sigma} (-1)^j \, 
  ( c_{j,\sigma}^\dagger c_{j+1,\sigma}+ \mathrm{H.c.} )
.
\end{equation}
Without loss of generality we can assume $\delta>0$.
When $V=0$, the model is called ``Peierls-Hubbard model.''
The one-dimensional Mott insulator, realized when
   $U>0$ and $V=0$, is known to be 
   unstable against the Peierls distortion, \cite{Bray,Ukrainskii} and
   as a result the ground state changes from the SDW state
   into the BCDW state regardless of the magnitude of the Hubbard
   interaction $U$.
Such an instability comes from the fact that
   the bond dimerization tends to concentrate the electron density
   onto bonds, without any conflict with
   the Hubbard, $U$, repulsion.\cite{Brune}
However, the nearest-neighbor Coulomb repulsion $V$ competes with
   this $\delta$ term, since the $V$ interaction likes to localize two
   electrons on a single site and promotes the CDW state.
Here we investigate the instability of the BCDW state against the
   intersite Coulomb repulsion $V$, and clarify the 
   critical behavior near the transition between the BCDW state and
   the CDW state.

The bond dimerization $H_\delta$ is bosonized as
   $H_\delta=\int dx\mathcal{H}_\delta$, where
\begin{equation}
\mathcal{H}_\delta =
 -\frac{g_\delta}{2(\pi a)^2} \cos \theta \, \cos \phi
\end{equation}
   and $g_\delta=8\pi \delta a$.
One finds that the EHM with the bond dimerization
   also has a two-component DSG structure.
Here
   the charge phase field $\theta$ is subjected to the potential
   $\cos \theta$ instead of $\sin \theta$ of the $g_\Delta$ term
   [Eq.\ (\ref{eq:Hw_bosonization})], while
   the locking potential for the spin phase field $\phi$ has 
   the same structure as that of the $g_\Delta$ term.

It is important to note that the BCDW order parameter
   $\mathcal{O}_\mathrm{BCDW}$ takes a nonvanishing expectation value
   for any $U$ and $V$ if $\delta \ne0$,
   as $\mathcal{H}_\delta\propto\mathcal{O}_\mathrm{BCDW}$.
In this section we will not use the term BCDW to characterize
   phases, and, in particular, the phase containing the trivial
   Peierls insulator ($U=V=0$ and $\delta\ne0$) is called the Peierls
   insulating (PI) phase.

\begin{table}[t]
\caption{Possible ground states  and
  the position of locked phase fields,
  determined from Eq.\ (\protect{\ref{eq:VD0}}).
}
\label{table:phase-locking-D}
\begin{ruledtabular}
\begin{tabular}{lc}
Phase  &  $(\theta,\phi )$       \\
\hline
SDW   &  $ (0,\pm\gamma_\phi ), 
                (\pi,\pm(\pi-\gamma_\phi) )$  \\
CDW   &  $ (\pm \gamma_\theta, 0 ),
                (\pm(\pi - \gamma_\theta), \pi )$  \\
PI (for $g_\delta>0$)   &  
              $(0,0 ), (\pi,\pi )$  \\
PI (for $g_\delta<0$)   &
              $(0,\pi ),(\pi,0 )$  \\
BSDW &        $(\pm \pi/2, \pm \pi/2 )$ 
\end{tabular}
\end{ruledtabular}
\end{table}
\subsection{Semiclassical analysis}

We begin with semiclassical analysis of the model with the $g_\delta$
   term.
We neglect spatial variations of the phase fields
   in $\mathcal{H}+\mathcal{H}_\delta$ and consider the potential
\begin{eqnarray}
V_\delta(\theta,\phi)
&=&
-g_c \cos2\theta + g_s \, \cos 2\phi
-g_{cs} \, \cos 2\theta \, \cos 2\phi 
\nonumber \\ 
&& {}
-g_\delta \, \cos\theta \, \cos \phi ,
\label{eq:VD}
\end{eqnarray}
   where $g_{cs}=g_{3\parallel}<0$.

First, we consider the simpler case where $g_{cs}=0$, which corresponds
   to the situation where $g_{cs}$ is irrelevant in the RG sense.
The potential in this case is
\begin{equation}
V_\delta^0(\theta,\phi) =
-g_c \cos2\theta + g_s \, \cos 2\phi - g_\delta \, \cos\theta \, \cos \phi.
\label{eq:VD0}
\end{equation}
The positions of the potential minima are determined by
   the saddle-point equations,
   $\partial V_\delta^0(\theta,\phi)/\partial \theta=0$ and  
   $\partial V_\delta^0(\theta,\phi)/\partial \phi=0$.
We find that the potential has
   the double-well structure for the $\theta$ ($\phi$) phase field
   when $g_c<-|g_\delta|/4$ ($g_s>|g_\delta|/4$). 
Here we introduce $\gamma_\theta^0$ and $\gamma_\phi^0$
   ($0 \le \gamma_\theta^0, \gamma_\phi^0 \le \pi$) defined by
\begin{equation}
\gamma_\theta^0 
  = \left| \cos^{-1} \left(-\frac{g_\delta}{4g_c}\right) \right|,
\quad
\gamma_\phi^0
  = \left| \cos^{-1} \left(\frac{g_\delta}{4g_s}\right) \right|
\end{equation}
   for $|g_\delta/g_c| \le 4$ and $|g_\delta/g_c|\le4$, respectively.
The solutions to the saddle-point equations can be classified
   into the following four classes:
(i) the PI state,
   $(\theta,\phi)=(0,0), (0,\pi ), (\pi,0)$, or $(\pi,\pi)$
[for $g_\delta>0$, the phase fields are locked at
   $(\theta,\phi)=(0,0)$ or 
   $(\pi,\pi)$, while for $g_\delta<0$ the phase fields are 
   locked as $(\theta,\phi)=(0,\pi)$ or $(\pi,0)$];
(ii) the pure BSDW state,
   $(\theta,\phi)=(\pi/2,\pm \pi/2)$ or $(-\pi/2,\pm \pi/2)$;
(iii) the ``SDW'' state with both the SDW order and the BCDW order,
   $(\theta,\phi)=(0,\pm \gamma_\phi^0)$ or
     $\biglb(\pi, \pm (\pi - \gamma_\phi^0)\bigrb)$; and
(iv) finally, the ``CDW'' state with both the CDW order and
   the BCDW order,
    $(\theta,\phi)=(\pm \gamma_\theta^0, 0 )$ or
    $\biglb(\pm(\pi - \gamma_\theta^0), \pi \bigrb)$.
\begin{figure}[t]
\includegraphics[width=7.cm]{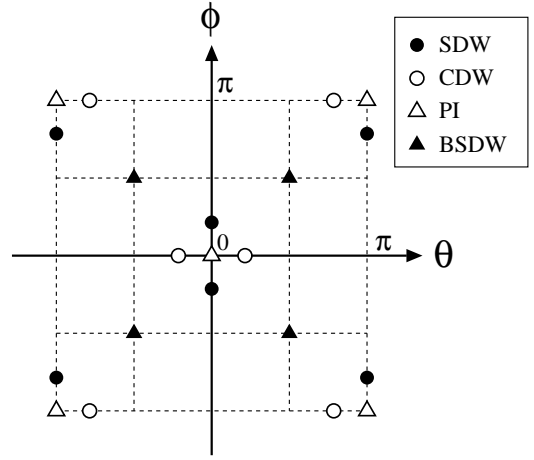}
\caption{
Positions of locked phase fields $\theta$ and $\phi$
   in the respective states for $g_\delta>0$. 
}
\label{fig:potmin-D}
\end{figure}%
The possible ground states and positions of locked phase fields
  are summarized in Table \ref{table:phase-locking-D} and
  Fig.\ \ref{fig:potmin-D}.
In these states the potential energy reads
\begin{subequations}
\begin{eqnarray}
V_{\rm PI}^0  &=&  - g_c + g_s  -|g_\delta| ,
\\
V_{\rm BSDW}^0  &=& + g_c -g_s ,
\\
V_{\rm SDW}^0
   &=&  - g_c - g_s - \frac{g_{\delta}^2}{8g_s} ,
\label{eq:V_SDW+BCDW}
\\
V_{\rm CDW}^0
   &=&  + g_c + g_s + \frac{g_{\delta}^2}{8g_c}.
\label{eq:V_CDW+BCDW}
\end{eqnarray}
\end{subequations}
In deriving Eqs.\ (\ref{eq:V_SDW+BCDW}) and 
   (\ref{eq:V_CDW+BCDW}), we have assumed $|g_\delta/g_s| \le 4$ and 
   $|g_\delta/g_c| \le 4$, respectively.
The PI state is stabilized by the first-order contribution 
   of the $g_\delta$ term.
Furthermore, if $g_s>0$ ($g_c<0$),
   the SDW state (the CDW state) is also stabilized due to 
   second-order contribution of $g_\delta$.
\begin{figure}[t]
\includegraphics[width=7cm]{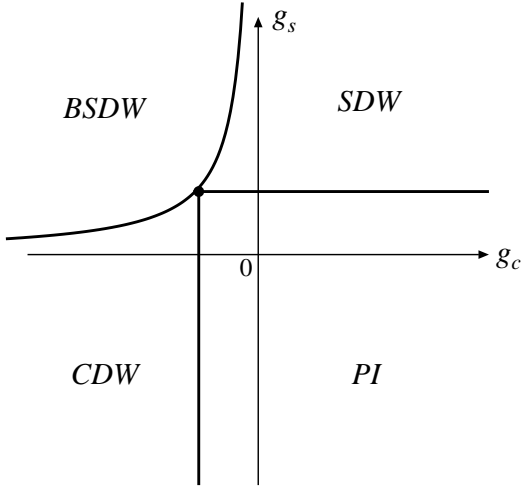}
\caption{
Phase diagram obtained by minimizing the potential energy 
   $V_\delta^0(\theta,\phi)$ [Eq.\ \protect{(\ref{eq:VD0})}].
The phase boundary of the BSDW state is given  by
  the curve $g_c g_s=-g_\delta^2/16$ with $g_c<0$.
The phase boundary between the PI state and the SDW state
   and that between the PI state and the CDW state are
   given by the lines $g_s= |g_\delta|/4$ with $g_c > |g_\delta|/4$
   and $g_c=-|g_\delta|/4$ with $g_s < |g_\delta|/4$, respectively.
All the phase transitions in this figure are continuous.
A multicritical point is at $(g_c,g_s)=(-|g_\delta|/4,|g_\delta|/4)$.
}
\label{fig:classical-D}
\end{figure}
\begin{figure}[t]
\includegraphics[width=7cm]{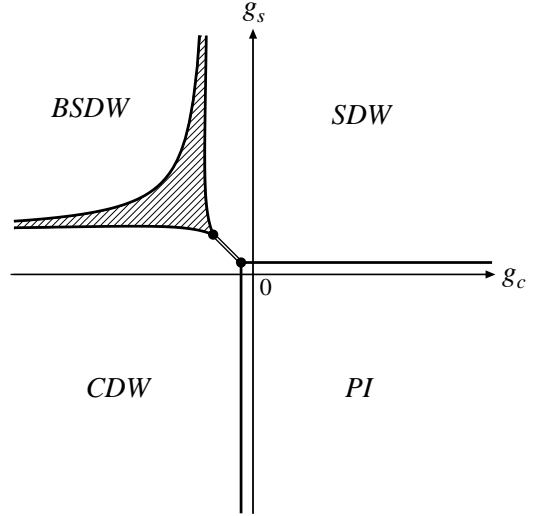}
\caption{
Phase diagram obtained by minimizing the potential energy 
   $V_\delta(\theta,\phi)$ [Eq.\ \protect{(\ref{eq:VD})}]
   drawn for $|g_\delta|/8<|g_{cs}|<|g_\delta|/4$.
Multicritical points are located at
   $(g_c,g_s)=(-|g_{cs}|,|g_{cs}|)$ and
   $(-\frac{1}{4}|g_\delta|+|g_{cs}|,\frac{1}{4}|g_\delta|-|g_{cs}|)$.
The boundary of the BSDW phase is
   $(g_c+|g_{cs}|)(g_s-|g_{cs}|)<-g_\delta^2/16$.
The edges of the PI phase are defined by the lines
   $g_c=-\frac{1}{4}|g_\delta|+|g_{cs}|$ and
   $g_s=\frac{1}{4}|g_\delta|-|g_{cs}|$.
The double line denotes a first-order transition, and the single lines
   denote continuous transitions.
Within the semiclassical analysis the ground state in the shaded
   region has the coexisting order of the SDW, CDW, BCDW, and BSDW.
}
\label{fig:classical-Dcs}
\end{figure}
The phase diagram obtained by comparing these energies 
   is shown in Fig.\ \ref{fig:classical-D}.

From the above semiclassical analysis one might conclude that 
   the topological charge $S_z$ [Eq.\ (\ref{eq:topological_charge})]
   becomes fractional in the SDW phase and that
   the Ising-type phase transition in the spin sector takes place
   on the boundary between the PI state and the SDW state.
However, as discussed in Sec.\ \ref{sec:staggered}, 
   the global SU(2) symmetry prohibits 
   the Ising criticality in the spin sector
   and changes
   the SDW phase in Fig.\ \ref{fig:classical-D}
   into the PI phase.

Next we include the $g_{cs}$ term.
Table \ref{table:phase-locking-D} still stands if we replace
$g_c$ and $g_s$ with $g_c-|g_{cs}|$ and $g_s+|g_{cs}|$ in
$\gamma_\theta^0$ and $\gamma_\phi^0$, respectively.
The phase diagram obtained by minimizing the potential energy
$V_\delta(\theta,\phi)$ is shown in Fig.~\ref{fig:classical-Dcs}.
New features compared with Fig.~\ref{fig:classical-D} are the appearance
of a first-order transition line
and of the new phase in which the ground state has the coexisting order
of the SDW, CDW, BCDW, and BSDW.
The new phase is shown as the shaded region in
Fig.~\ref{fig:classical-Dcs}, which is surrounded by the three curves
defined by
\begin{subequations}
\begin{eqnarray}
&&(g_c+|g_{cs}|)(g_s-|g_{cs}|)=-\frac{g_\delta^2}{16},\\
&&(g_c+|g_{cs}|)(g_s+|g_{cs}|)^2=-\frac{g_\delta^2}{16}(g_s-|g_{cs}|),\\
&&(g_s-|g_{cs}|)(g_c-|g_{cs}|)^2=-\frac{g_\delta^2}{16}(g_c+|g_{cs}|).
\qquad
\end{eqnarray}
\end{subequations}

Let us focus on the phases which can be realized when
   $g_s \simeq g_c$, in view of the fact that
   in the extended Hubbard model both
   $g_s(=g_{1\perp})$ and $g_c(=g_{3\perp})$ are given by 
   $(U-2V)$ in the lowest order.
Along the line $g_s\simeq g_c$ in Figs.\ \ref{fig:classical-D}
   and \ref{fig:classical-Dcs}, 
   there are three possible phases: the SDW state, 
   the PI state, and the CDW state.
Since the SDW state is prohibited by the SU(2) symmetry
   and becomes the PI state, we expect to have only two phases,
   the PI state and the CDW state, and a single phase
   transition between them.
The transition is continuous at $|g_{cs}/g_\delta|\ll1$ and changes into
a discontinuous transition when $g_{cs}$ exceeds $|g_\delta|/4$.

\subsection{Renormalization-group analysis}

Next we perform perturbative RG analysis to take into account quantum
   fluctuations.
The one-loop RG equations for coupling constants
   in $\mathcal{H}+\mathcal{H}_\delta$ are given by
\begin{eqnarray}
\frac{d}{dl} G_\delta
  &=& {} + G_\delta
    + \frac{1}{2} \, G_\delta \, G_\rho 
    + G_\delta \, G_c
\nonumber \\ && {}
    - \frac{3}{2} \, G_\delta \, G_s
    + \frac{3}{4} \, G_\delta \, G_{cs}
    - \frac{3}{8} \, G_\delta \, G_{\rho s}
,
\label{eq:RGd-b}
\\
\frac{d}{dl} G_\rho 
  &=& {} + \frac{1}{4} \, G_\delta^2
    + 2 \, G_c^2 + G_{cs}^2 +  G_s \, G_{\rho s}
,
\\
\frac{d}{dl} G_c 
  &=& {}
  + \frac{1}{4} \, G_\delta^2 
    + 2 \, G_\rho\, G_c
  - G_s \, G_{cs} - G_{cs} \, G_{\rho s}
, \nonumber \\
\\
\frac{d}{dl} G_s 
  &=& {} - \frac{1}{4} \, G_\delta^2 
    - 2 \, G_s^2 - G_c \, G_{cs} - G_{cs}^2
,
\\
\frac{d}{dl} G_{cs} 
  &=& {} + \frac{1}{4}  \,G_\delta^2 
    - 2 \, G_{cs} + 2 \, G_\rho \, G_{cs} - 4 \, G_s \, G_{cs}
\nonumber \\ && {}
    - 2 \, G_c\, G_{s}
    - 2 \, G_c\, G_{\rho s}
    - 4 \, G_{cs} \, G_{\rho s}
,
\\
\frac{d}{dl} G_{\rho s} 
  &=& {} - \frac{1}{4}  \, G_\delta^2 
    - 2 \, G_{\rho s} + 2 \, G_\rho \, G_s
\nonumber \\ && {}
    - 4 \, G_c \, G_{cs} - 4 \, G_{cs}^2
    - 4 \, G_s \, G_{\rho s}
.
\label{eq:RGd-e}
\end{eqnarray}
The initial value of $G_\delta(l)$ is given by 
   $G_\delta(0)=2\delta/t$ and those of the other coupling constants
   are $G_\nu(0)=g_\nu /(4\pi t)$.
We note that these RG equations are invariant under 
   the sign change of $G_\delta(l)$.
We can thus assume $G_\delta(0) \ge 0$ without losing generality.

To find the ground-state phase diagram of the system,
   we solve the scaling equations 
   (\ref{eq:RGd-b})--(\ref{eq:RGd-e}) numerically,
   as in the preceding sections.
We determine to which phase the ground state belongs by
   looking at which one of the couplings
   $G_\delta(l)$, $G_c(l)$, $G_s(l)$, and $G_{cs}(l)$ becomes most relevant. 
For repulsive $U$ and $V$ there are four possibilities as listed below.

(i)
If $G_c$ is most relevant and $G_c(l) \to 1 $ at $l=l_{\rho+}$,
   then
   the phase field $\theta$ is locked at $\theta=0$ or $\pi$, 
   and 
   the effective Hamiltonian for the spin sector at $l\ge l_{\rho+}$
   becomes
\begin{eqnarray}
\mathcal{H}^{\mathrm{eff}}_{\sigma+}
&=&
\frac{v_F}{2\pi} \sum_{p=\pm} (\partial_x \phi_p)^2
 - \frac{v_F}{\pi} \, G_s^* (\partial_x \phi_+ )(\partial_x \phi_-) 
\nonumber \\ && {}
     \mp \frac{v_F}{\pi a^2} \, G_\delta^* \,  \cos \phi
     +  \frac{v_F}{\pi a^2} \, G_s^* \, \cos 2\phi,
\end{eqnarray}
   where $G_s^*=G_s(l_{\rho+})-G_{cs}(l_{\rho+})$ and 
   $G_\delta^*=G_\delta(l_{\rho+})$,
  and the sign $-/+$ of the $G_\delta^*$ term
   corresponds to the location of the phase locking $\theta=0/\pi$.
This effective theory is the same as Eq.\ (\ref{eq:Heff_sigma-}).
As seen before, 
   regardless of the sign of $G^*_s$, 
   the phase field $\phi$ is locked at $\phi=0$ or $\pi$
   depending on the position of the charge phase locking 
   $\theta=0$ or $\pi$.
Thus we have the phase locking $(\theta,\phi)
    =(0,0)$ or $(\pi,\pi)$, 
   i.e., the PI state as the ground state.
We note that due to the SU(2) spin rotation symmetry
   the SDW state cannot be realized even if $G_s^*>0$.

(ii)
If $G_c$ is most relevant and $G_c(l) \to -1$ at $l=l_{\rho-}$,
   then the phase field $\theta$ is locked at 
   $\theta = \pm \pi/2$.
The effective Hamiltonian for the spin part is
\begin{eqnarray}
\mathcal{H}^{\mathrm{eff}}_{\sigma -}
&=&
\frac{v_F}{2\pi} \sum_p (\partial_x \phi_p)^2
     -  \frac{v_F}{\pi} \, G_s^*    (\partial_x \phi_+ )(\partial_x \phi_-) 
\nonumber \\ && {}
     + \frac{v_F}{\pi a^2} \, G_s^* \, \cos 2\phi,
\end{eqnarray}
   where $G_s^* = G_s(l_{\rho-}) + G_{cs}(l_{\rho-})$.
We have verified numerically that $G_s^*$ always becomes negative
   in this case.
The $G_s^*$ terms are then marginally relevant
   [$G_s^*(l)\to - \infty$ under scaling].
The phase fields are then locked at 
   $( \theta , \phi )= ( \pm \pi/2, 0), ( \pm \pi/2, \pi )$,
   which corresponds to the CDW phase
   with $\gamma_\theta \to \pi/2$
   (i.e., $g_\delta \to 0$, see Table \ref{table:phase-locking-D}).
Since $\mathcal{H}_\delta\propto\mathcal{O}_\mathrm{BCDW}$,
   the order parameter of
   the BCDW should have a nonvanishing expectation value.
We thus conclude that the ground state is in the CDW phase.

(iii)
If either $G_\delta$ or $G_{cs}$ is most relevant,
   both charge and spin fluctuations are suppressed.
In this case the semiclassical treatment is justified, and
   we can determine to which phase the ground state belongs
   by substituting $G_c$ and $G_s$ to $g_c$ and $g_s$
   in Fig.\ \ref{fig:classical-Dcs}.

(iv)
If $G_s$ is most relevant and $G_{s}(l)\to -1$ at $l=l_\sigma$,
   the spin fluctuations are suppressed and  
   the phase field $\phi$ is locked at $\phi \to 0$ or $\pi$ 
   below this energy scale.
The effective Hamiltonian at $l\ge l_\sigma$ is given by
\begin{eqnarray}
\mathcal{H}^{\mathrm{eff}}_{\rho}
&=&
\frac{v_F}{2\pi} \sum_{p=\pm} (\partial_x \theta_p)^2
+ \frac{v_F}{\pi} \, G_\rho^*
  \left(\partial_x \theta_+ \right)
  \left(\partial_x \theta_- \right)
\nonumber \\ && {}
\mp \frac{v_F}{\pi a^2} \, G_\delta^* \, \cos \theta
-\frac{v_F}{\pi a^2} \, G_c^* \, \cos 2\theta, 
\end{eqnarray}
   where $G_\rho^*=G_\rho(l_\sigma)-G_{\rho s}(l_\sigma)$,
   $G_c^*=G_c(l_\sigma)+G_{cs}(l_\sigma)$, and  
   $G_\delta^*=G_\delta(l_\sigma)$.
The sign $-/+$ of the $G_\delta^*$ term 
   corresponds to the phase locking $\phi=0/\pi$.
Both of the nonlinear terms $\cos \theta$ and
   $\cos 2\theta$ are relevant perturbations.
If $G_c^*<0$,
   these two terms compete with each other, and
   this DSG model exhibits the Ising criticality.
The ground state is either in the PI phase or
   in the CDW phase, and there is an Ising-type quantum phase
   transition between the two phases.
Here we estimate the Ising critical point
   from the semiclassical analysis.
That is, the critical value is determined from the condition
   $G_c^*/G_\delta^*=-1/4$ (see Fig.\ \ref{fig:classical-D}). 
If $G_c^*>0$, these two terms do not compete and thus the
   phase locking is $\theta=0$ $(\pi)$ for $\phi=0$ $(\pi)$,
   where the ground state is the PI state.

\begin{figure}[t]
\includegraphics[width=7cm]{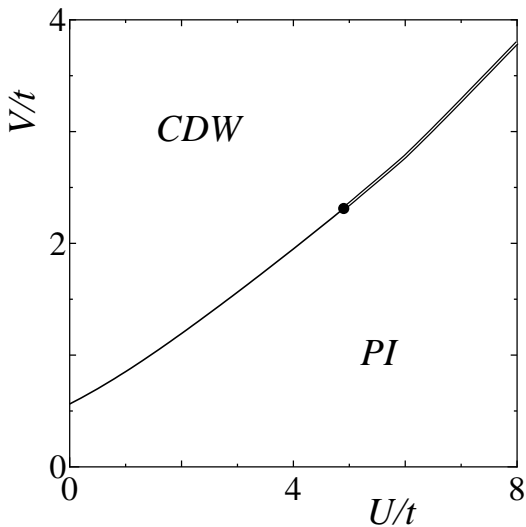}
\caption{
Phase diagram 
   of the half-filled extended Hubbard model with
   $\delta/t=0.1$.
The second-order transition line (single line) turns into
   the first-order transition line (double line) at the
   tricritical point $(U_c,V_c)\approx(4.9t, 2.3t)$.
}
\label{fig:UVD}
\end{figure}

The resultant phase diagram
  in the $U$-$V$ plane is shown in Fig.\ \ref{fig:UVD}.
In the weak-coupling region,
  the transition from the PI state to the 
   CDW state is characterized by the appearance of the
   double-well structure of the effective potential to the
   $\theta$ field, and
   thus the phase transition in Fig.\ \ref{fig:UVD} belongs to the
   Ising universality class.
As we increase $U$ and $V$,
    there appears a tricritical point at $(U_c,V_c)\approx (4.9t,2.3t)$,
    where the phase transition 
    changes from second order to first order.

\begin{figure}[b]
\includegraphics[width=7.5cm]{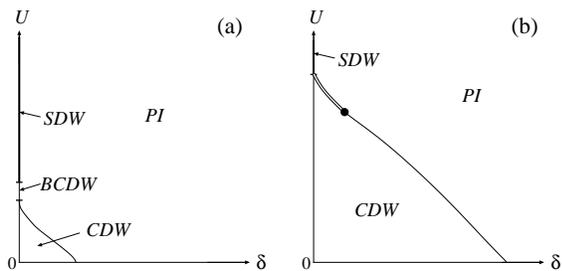}
\caption{
Schematic phase diagram of the half-filled extended Hubbard model
 at (a) $V\ll t$ and (b) $V\gg t$.
The single lines represent second-order transitions, and the
 double line represents a first-order transition.
}
\label{fig:schematicDU}
\end{figure}

\begin{figure}[t]
\includegraphics[width=8.5cm]{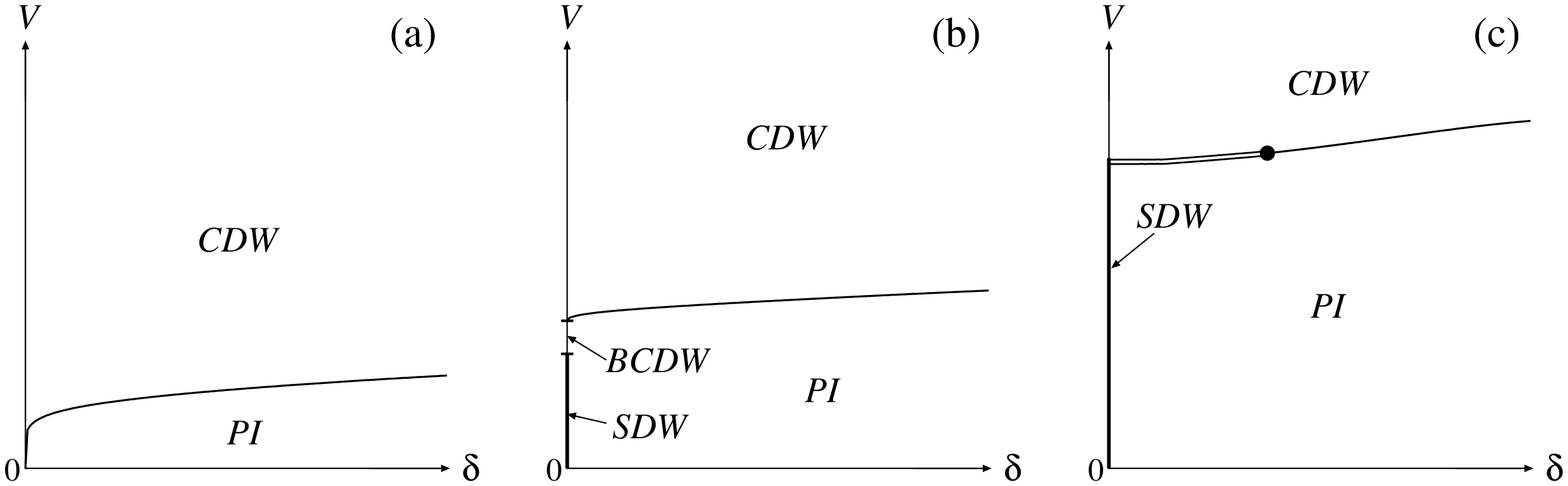}
\caption{
Schematic phase diagram of the half-filled extended Hubbard model
 at (a) $U=0$, (b) $U\ll t$, and (c) $U\gg t$.
The single lines represent second-order transitions, and the
 double line represents a first-order transition.
}
\label{fig:schematicDV}
\end{figure}

Figure \ref{fig:schematicDU} shows schematic phase diagrams in the
$\delta$-$U$ plane for $V\ll t$ and $V/t\gg1$.
When $\delta=0$, 
  we obtain three phases (the CDW, BCDW, and SDW phases) for $V\ll t$ (a) 
   and two phases (the CDW and SDW phases) for $V\gg t$ (b), 
  as we discussed
   in Sec.\ \ref{sec:phase_diagram} (see Fig.\ \ref{fig:phase2}).
Upon turning on $\delta$, the SDW ground state changes into the PI state,
   where the transition is described by the Gaussian theory.
On the other hand, the BCDW state changes into the PI state without
   accompanying any singularity: This change is merely lifting of the
   doubly degenerate  BCDW ground states.

Figure \ref{fig:schematicDV} shows schematic phase diagrams in
the $\delta$-$V$ plane for $U=0$,  $U/t\ll1$, and $U/t\gg1$.
At $U=0$ we have a single critical value $V_c$ 
   which has the $\delta$ dependence given by
   $V_c \propto 1/\ln(t/\delta)$ for small $\delta$.  
As $U$ and $V$ increase, the phase boundary approaches
   the $U=2V$ line.
The asymptotic form of $V_c$ for $U,V\gg \delta$
   and $\delta\ll t$ 
is given by $V_c=\frac{1}{2}U+ C'' \, U (\delta/t)^{2U/\pi t}$,
   where $C''$ is a numerical constant of the order of unity
  (see also Fig.\ \ref{fig:UVD}).

\section{Conclusions}\label{sec:conclusions}

In this paper we have 
   studied the ground-state phase diagram of the 
   one-dimensional extended Hubbard model with 
   on-site and nearest-neighbor repulsion $U$ and $V$.
By including higher-order corrections to coupling constants in 
   the $g$-ology,
   we have given a plausible theoretical argument within the RG
   approach for the mechanism of the appearance of the BCDW phase
   at $U\approx 2V$ in the weak-coupling limit.
Our two-step RG approach, however, is not complete in that there
   remains a weak cutoff dependence in the phase boundaries.
This, albeit minor, defect should be resolved with use of a more
   sophisticated systematic RG procedure.
Away from the weak-coupling limit the umklapp scattering between
   the parallel-spin electrons $g_{3\parallel}$ tends to destabilize 
   the BCDW state and eventually gives rise to a bicritical point
   where the two continuous-transition lines merge
   into the SDW-CDW first-order transition line 
   (Fig.\ \ref{fig:phase2}).
We should note, however, that there still remains a difficult
   question as to whether our phase diagram is qualitatively correct
   near the multicritical point (which we call bicritical).
One could imagine, for example, a possibility that a continuous phase
   transition between the BCDW state and the CDW state becomes first
   order before reaching the multicritical point, due to higher-order
   effects that are ignored in our analysis.
If the correct topology of the phase diagram is indeed the same as
   ours (Fig.\ \ref{fig:phase2}), then the critical properties of
   the multicritical point remain to be understood.
We hope that these issues will be resolved by future studies.

We have also examined effects of additional staggered site potential
   and bond dimerization in the extended Hubbard model.
In the presence of the staggered site potential,
   we have found that the BCDW state is smoothly connected to the
   SDI phase which is obtained for $V=0$ by
   Fabrizio \textit{et al}. \cite{Fabrizio} 
In this BCDW phase the BCDW order coexists with the CDW order, 
   and the quantum phase transition
   between the BI phase (or the CDW phase) and
   the BCDW phase belongs to the Ising universality class
   ($c=\frac{1}{2}$ CFT).
For finite $V$ the BCDW phase is also destabilized by the
   $g_{3\parallel}$ term, and the direct first-order quantum phase
   transition between the SDW state (= Mott insulating state) and
   the BI state takes place (Fig.\ \ref{fig:UVW}).
In the presence of the staggered bond dimerization
   the SDW phase becomes unstable and the ground state at $V=0$ turns
   out to be the Peierls insulating state.
For $V\ne0$ the phase diagram consists of two phases,
   the PI state and the CDW state, which are
   separated by a phase transition line of the Ising criticality
   (Fig.\ \ref{fig:UVD}).

\acknowledgments

One of the authors (M.T.) thanks
    E.\ Orignac, M.\ Sugiura, 
   Y.\ Suzumura, K.\ Yonemitsu, and H.\ Yoshioka 
   for valuable discussions.
The authors also thank  S.\ Qin for useful discussions.
A.F.\ is grateful to Aspen Center for Physics for its hospitality,
   where this paper was finished.
This work was supported in part by a Grant-in-Aid for
   Scientific Research on Priority Areas from the Ministry of
   Education, Culture, Sports, Science and Technology, Japan
   (Grant No.\ 12046238).

\appendix

\section{Bosonization}\label{sec:bosonization}

In this section, we derive the phase Hamiltonian of the
   1D extended Hubbard model by using
   the Abelian bosonization method. \cite{Shankar}
We include not only the marginal terms but the
   leading irrelevant terms which play a crucial role
   in the first-order SDW-CDW transition at strong coupling.

The Lagrangian for the free massless boson theory 
   in a two-dimensional Euclidean
   space is given by
\begin{eqnarray}
L_\theta &=&
\frac{1}{4\pi} \int dx 
\left[
v \left(\partial_x \theta \right)^2 + \frac{1}{v}
   \left(\partial_\tau \theta \right)^2
\right]
,
\end{eqnarray}
   where $\theta$ is a bosonic field, $\tau$ is the imaginary time,
   and $v$ is velocity.
The variable canonically conjugate to $\theta$ is given by
\begin{eqnarray}
\Pi &\equiv& \frac{\partial L}{\partial \dot{\theta}}
=\frac{i}{2\pi v} \partial_\tau \theta
,
\end{eqnarray}
   where $\dot{\theta}=\partial \theta/ \partial t$ and
   $t$ is the real time ($\tau = it$).
As usual this system is quantized by imposing
   the commutation relation at equal times:
  $[\theta(x), \Pi(x')] = i\delta(x-x')$.
Thus the Hamiltonian for the free boson theory is given by
 $H_\theta = i\int dx \, \Pi \, \partial_\tau \theta + L_\theta$, i.e.,
\begin{equation}
H_\theta =
\frac{v}{4\pi} \int dx 
\left[
  \left(2\pi \Pi\right)^2 +
 \left(\frac{d\theta}{dx} \right)^2 
\right]
.
\end{equation}

Introducing two copies of this theory with fields
   $\theta$ and $\phi$ and velocity $v=v_F$,
   we arrive at $H_0$ [Eq.\ (\ref{eq:H0})], where
   the fields $\theta$ and $\phi$ represent
   the ``charge'' and ``spin'' degrees of freedom.
The chiral bosonic fields $\theta_\pm(x,\tau)$ and
   $\phi_\pm(x,\tau)$
   are introduced in Eqs.\ (\ref{eq:chiral_theta}) and 
   (\ref{eq:chiral_phi}), respectively, where
   the right-moving (left-moving) fields 
   are functions of 
   $\tau-i(x/v_F)$ [$\tau+i(x/v_F)$].\cite{Shankar}  
The phase field $\theta$ ($\phi$) and its dual phase field
   $\tilde{\theta}$ ($\tilde{\phi}$) are written in terms of
   the chiral fields as
\begin{eqnarray}
&&
\theta = \theta_+ + \theta_-
,
\hspace*{0.2cm}
\tilde{\theta} =\theta_+ - \theta_-
, 
\\
&&
\phi = \phi_+ + \phi_-
,
\hspace*{0.2cm}
\tilde{\phi} = \phi_+ - \phi_-
.
\end{eqnarray}
They satisfy the following commutation relations:
\begin{eqnarray}
[ \theta(x),\tilde{\theta}(x') ] 
=
 [\phi (x), \tilde{\phi} (x')] 
= -i \, 2\pi \, \Theta(-x+x')
, 
\end{eqnarray}
where $\Theta(x)$ is the Heaviside step function.

The electron field operators $\psi_{p,\sigma}(x)$ are given in
   Eq.\ (\ref{eq:field_op}) in terms of
   a new set of chiral bosonic fields
   $\varphi_{p,\sigma}$ introduced in Eq.\ (\ref{eq:varphi}).
In this bosonization scheme $\psi_{+,\sigma}$ and $\psi_{-,\sigma}$
   anticommute, and we only need to introduce the Klein factor
   $\kappa_\sigma$ to ensure the anticommutation relation between
   fields with different spins;
   cf.\ the so-called \textit{constructive} bosonization
   method.\cite{vonDelft}
From Eqs.\ (\ref{eq:commutation_varphi}) and (\ref{eq:field_op})
   the electron-density operator becomes
\begin{equation}
\rho_{p,\sigma}(x)
\equiv
{} : \psi_{p,\sigma}^\dagger \, \psi_{p,\sigma}\!: {}
=
\frac{1}{2\pi} \, \frac{d}{dx} \, \varphi_{p,\sigma}(x) 
.
\label{eq:density}
\end{equation}
As is well known, the Hamiltonian density of free bosons (\ref{eq:H0}),
  i.e., 
\begin{eqnarray}
\mathcal{H}_{0}
=
\frac{v_F}{4\pi} \sum_{p=\pm} \sum_{\sigma}
    \left( \frac{d\varphi_{p,\sigma}}{dx} \right)^2
=
\pi v_F \sum_{p,\sigma} 
\rho^2_{p,\sigma}(x)
,
\end{eqnarray}
is equivalent to the Hamiltonian density of free fermions with linear
energy dispersion, Eq.\ (\ref{eq:H0_linear}).
This can be shown, for example, by using the OPE method.\cite{Affleck}

Next we bosonize the interaction term $H_\mathrm{int}$.
Without the nearest-neighbor repulsion $V$, 
   this can be easily done as\cite{Solyom,Emery}
\begin{eqnarray}
\mathcal{H}_{\rm int}^{V=0}
&=&
\frac{g_{4\parallel}+g_{4\perp}}{4\pi^2} 
\left[ \left(\partial_x \theta_+ \right)^2 +
       \left(\partial_x \theta_- \right)^2  \right]
\nonumber \\ && {}
+\frac{g_{4\parallel}-g_{4\perp}}{4\pi^2} 
\left[ \left(\partial_x \phi_+ \right)^2 +
       \left(\partial_x \phi_- \right)^2  \right]
\nonumber \\ && {}
+\frac{g_{2\parallel}+g_{2\perp}-g_{1\parallel}}{2\pi^2} 
   \bigl(\partial_x \theta_+ \bigr)
       \bigl(\partial_x \theta_- \bigr) 
\nonumber \\ && {}
+\frac{g_{2\parallel}-g_{2\perp}-g_{1\parallel}}{2\pi^2} 
   \bigl(\partial_x \phi_+ \bigr)
       \bigl(\partial_x \phi_- \bigr) 
\nonumber \\ && {}
-\frac{g_{3\perp}}{2(\pi a)^2} 
   \cos 2\theta
+\frac{g_{1\perp}}{2(\pi a)^2} 
    \cos 2\phi
,
\label{eq:H_lowest}
\end{eqnarray}
   where $g$'s are given in and below Eq.\ (\ref{eq:g}).
In the presence of $V$, 
   the matrix element of the umklapp process with parallel spins
   $H_{g_{3\parallel}}$ [the $g_{3\parallel}$ process in 
   Eq.\ (\ref{eq:HI_fourier})] has a finite amplitude at lowest 
   order in $g$-ology.
This term can be bosonized as
\begin{eqnarray}
\mathcal{H}_{g_{3\parallel}}
&=& {}
-\frac{g_{3\parallel}}{2(\pi a)^2} \cos 2\theta \, \cos 2\phi
,
\label{eq:g3para}
\end{eqnarray}
   where $g_{3\parallel}=-2Va$ in the lowest order in $V$.
This term, which couples the charge and spin degrees of freedom,
   is often neglected since it is an irrelevant perturbation
   with scaling dimension $4$,
   consisting of $\mathrm{dim}[\cos2\theta]=2$ plus 
   $\mathrm{dim}[\cos2\phi]=2$.
Cannon and Fradkin were the first to suggest that this term should play
   an important role in the first-order SDW-CDW transition in the
   half-filled EHM. \cite{Cannon}
Voit then derived RG equations including this term.
However he did not include all the operators with scaling dimension 4
   and failed to keep the spin-rotational SU(2) symmetry.\cite{Voit}
We have to be careful in dealing with the $V$ interaction to
   include the important terms with scaling dimension up to 4.
To this end, we focus on the $V$ interaction and bosonize each
   scattering process separately.

First, the $g_{1\parallel}$ term\cite{Solyom} representing the backward
   scattering with parallel spins
   is bosonized by using Eq.\ (\ref{eq:field_op}) as
\begin{eqnarray}
&& \hspace*{-.7cm}
Va \sum_{p,\sigma} 
  \psi^\dagger_{p,\sigma}(x) \, \psi_{-p,\sigma}(x) \,
  \psi^\dagger_{-p,\sigma} (x+a) \, \psi_{p,\sigma}(x+a)
\nonumber \\
&=& -\frac{Va}{(2\pi a)^2} \sum_{p,s=\pm} 
  e^{ip[\theta(x+a)-\theta(x)]+ips [\phi(x+a)-\phi(x)]}
\nonumber \\
&=& 
\frac{Va}{2\pi^2} 
\left[ \sum_p \left(\partial_x \theta_p \right)^2
      + 2 \left(\partial_x \theta_+ \right)
          \left(\partial_x \theta_- \right) 
\right]
\nonumber \\ && {}
+\frac{Va}{2\pi^2} 
\left[  \sum_p \left(\partial_x \phi_p \right)^2 
      + 2 \left(\partial_x \phi_+ \right)
          \left(\partial_x \phi_- \right) 
\right]
\nonumber \\ && {}
-\frac{Va}{4\pi^2}  \, a^2 
\left(\partial_x \theta\right)^2 \left(\partial_x \phi\right)^2
+ \cdots ,
\label{eq:Hg1paraV}
\end{eqnarray}
   where we have expanded the exponent in the second line up to the
   order $a^4$
   for the $\theta$ sector and the $\phi$ sector, separately.
Since we are interested in operators that couple $\theta$ and $\phi$
   as in Eq.\ (\ref{eq:g3para}),
   we have discarded dimension-4 terms such as $a^4(\partial_x \theta)^4$
   and $a^4(\partial_x \phi)^4$ that involve only one sector.
Such terms as $(\partial_x\theta_+)(\partial_x\theta_-)$
   and $(\partial_x\phi_+)(\partial_x\phi_-)$ are
   already retained in Eq.\ (\ref{eq:H_lowest}), 
   while the last term proportional to 
   $(\partial_x\theta)^2 (\partial_x\phi)^2$ is a new term 
   with scaling dimension 
   $2+2$, which was missed in Ref.\ \onlinecite{Voit}.
We note that the Fermi velocity is renormalized by the $g_{1\parallel}$ term
   due to the presence of
   $\sum_p(\partial_x \theta_p)^2$ and  
   $\sum_p(\partial_x \phi_p)^2$.
This is in contrast with the conventional treatment where
   the velocity renormalization comes only from
   the forward scattering term $g_4$.\cite{Solyom}

In a similar way, the interaction terms of backward and umklapp 
   scattering with opposite spins 
   (so-called $g_{1\perp}$ and $g_{3\perp}$ terms,\cite{Solyom} 
    respectively)
   are bosonized as 
\begin{eqnarray}
&& \hspace*{-1.cm}
Va \sum_{p,\sigma}
  \psi^\dagger_{p,\sigma}(x) \, 
  \psi_{-p,\sigma}(x) \,
  \psi^\dagger_{-p,\overline{\sigma}} (x+a) \, 
  \psi_{p,\overline{\sigma}}(x+a)
\nonumber \\
&=& -\frac{Va}{(2\pi a)^2} \sum_{p,s=\pm} 
  e^{ip[\theta(x+a)-\theta(x)]-ips [\phi(x+a)+\phi(x)]}
\nonumber \\
&=& 
-\frac{2Va}{2(\pi a)^2} \cos 2\phi
+ \frac{2Va}{2\pi^2} 
   \left(\partial_x \theta_+ \right)
   \left(\partial_x \theta_- \right) \, 
   \cos 2\phi
\nonumber \\ && {}
+ \frac{2Va}{4\pi^2} 
  \left[ \sum_p \left(\partial_x \theta_p \right)^2 \right]
   \cos 2\phi 
+ \cdots,
\label{eq:Hg1perpV}
\\
&& \hspace*{-1.cm}
Va \sum_{p,\sigma} 
  \psi^\dagger_{p,\sigma}(x) \, 
  \psi_{-p,\sigma}(x) \,
  \psi^\dagger_{p,\overline{\sigma}} (x+a) \, 
  \psi_{-p,\overline{\sigma}}(x+a)
\nonumber \\
&=& \frac{Va}{(2\pi a)^2} \sum_{p,s=\pm} 
  e^{-ip[\theta(x+a)+\theta(x)]+ips [\phi(x+a)-\phi(x)]}
\nonumber \\
&=& 
+\frac{2Va}{2(\pi a)^2} \cos 2\theta
- \frac{2Va}{2\pi^2}
   \left(\partial_x \phi_+ \right)
   \left(\partial_x \phi_- \right) \,
   \cos 2\theta
\nonumber \\ && {}
- \frac{2Va}{4\pi^2} 
  \left[ \sum_p \left(\partial_x \phi_p \right)^2 \right]
   \cos 2\theta 
+ \cdots ,
\label{eq:Hg3perpV}
\end{eqnarray}
   where $\overline{\sigma}=\downarrow(\uparrow)$ for 
   $\sigma=\uparrow(\downarrow)$.
The potential $\cos 2\phi$ in Eq.\ (\ref{eq:Hg1perpV}) and 
   the potential $\cos 2\theta$ in Eq.\ (\ref{eq:Hg3perpV}) are
   already retained in Eq.\ (\ref{eq:H_lowest}), 
   while the other terms are new and have the scaling dimension $2+2$.

The forward-scattering terms
   ($g_{2\parallel}$, $g_{2\perp}$, $g_{4\parallel}$, and $g_{4\perp}$)
   do not generate operators of dimension $2+2$.

Hence the total Hamiltonian is given by
\begin{eqnarray}
\mathcal{H} &\!=\!&
\frac{1}{2\pi} 
     \sum_p 
   \left[ v_{\rho} \left(\partial_x \theta_p \right)^2
        + v_{\sigma} \left(\partial_x \phi_p \right)^2 \right]
\nonumber \\ && {}
+ \frac{g_\rho}{2\pi^2} 
     \left(\partial_x \theta_+ \right)
     \left(\partial_x \theta_- \right)
- \frac{g_\sigma}{2\pi^2} 
     \left(\partial_x \phi_+ \right) 
     \left(\partial_x \phi_- \right)
\nonumber \\ && {}
-\frac{g_{3\perp}}{2\pi^2 a^2} 
    \, \cos 2 \theta 
+\frac{g_{1\perp}}{2\pi^2 a^2}  
  \, \cos 2 \phi
\nonumber \\ && {}
+\frac{Va}{\pi^2 a^2} 
   \, \cos 2\theta  \,  \cos 2\phi
\nonumber \\ && {}
+\frac{Va}{2 \pi^2}
     \left[ \sum_{p}\left(\partial_x \theta_p \right)^2 
          + 2  \left(\partial_x \theta_+ \right) 
               \left(\partial_x \theta_- \right)  \right]
     \cos 2\phi
\nonumber \\ && {}
-\frac{Va}{2 \pi^2} 
     \left[ \sum_{p}\left(\partial_x \phi_p \right)^2 
          + 2 \left(\partial_x \phi_+\right)
              \left(\partial_x \phi_- \right) \right]
     \cos 2\theta 
\nonumber \\ && {}
-\frac{Va^3}{4\pi^2}  
     \left[ \sum_{p}\left(\partial_x \theta_p \right)^2 
          + 2  \left(\partial_x \theta_+ \right) 
               \left(\partial_x \theta_- \right)  \right]
\nonumber \\ && {} \quad\quad \times
     \left[ \sum_{p}\left(\partial_x \phi_p \right)^2 
          + 2 \left(\partial_x \phi_+\right)
              \left(\partial_x \phi_- \right) \right]
.
\label{eq:Hamiltonian_all}
\end{eqnarray}
The renormalized velocities are given by
  $v_\rho = 2ta + (U+6V) a/(2\pi)$ and
  $v_\sigma = 2ta - (U-2V) a/(2\pi)$.
The coupling constants $g_{1\perp}$ and $g_{3\perp}$ are defined in
   Eq.\ (\ref{eq:g}), and
   $g_\rho (\equiv g_{2\parallel}+g_{2\perp}-g_{1\parallel})$ and 
   $g_\sigma (\equiv -g_{2\parallel}+g_{2\perp}+g_{1\parallel})$
   are given by
\begin{subequations}
\begin{eqnarray}
g_\rho 
&=& 
(U+6V)a 
+ \frac{C_1}{4\pi t}(U-2V)^2a
+\frac{C_2}{\pi t} V^2 a
,
\\
g_\sigma
&=& 
(U-2V)a 
- \frac{C_1}{4\pi t}(U-2V)^2a
-\frac{C_2}{\pi t}V^2a
. \quad\quad\quad
\end{eqnarray} 
\end{subequations}
For the discussion of the SDW-CDW transition in the 1D EHM,
   it is sufficient to have the coupling constants of
   dimension 4 in lowest order in $V$.
We note that due to the SU(2) spin-rotation symmetry of the theory,
   the coupling constants for spin degrees of freedom must satisfy
   $g_{\sigma}=g_{1\perp}$, in any order of $U$ and $V$. 
To proceed further, we neglect the terms that involve
   $V\sum_p (\partial_x \theta_p )^2$ or
   $V\sum_p (\partial_x \phi_p )^2$ in Eq.\ (\ref{eq:Hamiltonian_all}).
These terms can lead to renormalization of the velocity through the
   RG transformation (see Appendix \ref{sec:rg}).
This effect can be ignored if we are interested in qualitative feature
   of the ground-state phase diagram of the model.
The final form of the bosonized Hamiltonian is thus given by
    Eq.\ (\ref{eq:Hamiltonian}).

\section{Derivation of renormalization-group equations}
\label{sec:rg}

In this section,
   we derive one-loop RG equations for the coupling constants including
   those operators with higher scaling dimension.
Our derivation is based on the operator product expansion (OPE) method.
The interaction part of the action $S_{\rm I}$ in the presence
   of the staggered site potential $G_\Delta$
   is given by 
\begin{eqnarray}
S_{\rm I} &=& {}
\frac{G_\rho}{\pi} \int d^2 r 
   \left(\partial_z \theta \right)
   \left(\partial_{\bar{z}} \theta \right) 
- \frac{G_\sigma}{\pi} \int d^2 r
   \left(\partial_z \phi \right) 
   \left(\partial_{\bar z} \phi \right) 
\nonumber \\ && {}
-\frac{G_\Delta}{\pi} \int \frac{d^2r}{a^2} \, 
  :\! \sin \theta \!: \, :\! \cos \phi \!:
\nonumber \\ && {}
-\frac{G_c}{\pi} \int \frac{d^2r}{a^2} \, :\! \cos 2\theta \!:
+\frac{G_s}{\pi} \int \frac{d^2r}{a^2} \, :\! \cos 2\phi \!:
\nonumber \\ && {}
-\frac{G_{cs}}{\pi} \int \frac{d^2r}{a^2} \, 
  :\! \cos 2\theta  \!: \, :\! \cos 2\phi \!:
\nonumber \\ && {}
-\frac{G_{\rho s}}{\pi} \int d^2r \,
   \left(\partial_z \theta \right) 
   \left(\partial_{\bar{z}} \theta \right) \,
   :\! \cos 2\phi \!:
\nonumber \\ && {}
+\frac{G_{c \sigma}}{\pi} \int d^2r \,
   \left(\partial_z \phi\right) 
   \left(\partial_{\bar z} \phi \right) \,
   :\! \cos 2\theta \!:
\nonumber \\ && {}
+\frac{G_{\rho \sigma}}{\pi} \int d^2r \, a^2 \,
   \left(\partial_z \theta \right) 
   \left(\partial_{\bar{z}} \theta \right) 
   \left(\partial_z \phi\right) 
   \left(\partial_{\bar z} \phi \right) ,
\end{eqnarray}
   where $z=v_F \tau+ix$, $\overline{z}=v_F \tau-ix$,
   $d^2r = v_F dx \, d\tau$, and $G_i=g_i/2\pi v_F$.
In this section,
   the operators are explicitly normal ordered.

In order to derive the RG equations, 
   we use the following  OPE's: 
\begin{subequations}
\begin{eqnarray}
&&
J_\rho (z) \, J_\rho (w)
=
\frac{1}{(z-w)^2} + \cdots ,
\\
&&
\bar{J}_\rho (\bar{z}) \, \bar{J}_\rho (\bar{w})
=
\frac{1}{(\bar{z}-\bar{w})^2} + \cdots ,
\\
&&
J_\rho (z)  :\! e^{i\alpha \theta(w,\bar{w})}\!:
=
\frac{\alpha}{2(z-w)} 
:\! e^{i\alpha \theta(w,\bar{w})}\!:
+ \cdots
, \,\,
\\
&&
J_\rho (\bar{z})  :\! e^{i\alpha \theta(w,\bar{w})}\!:
=
\frac{-\alpha}{2(\bar{z}-\bar{w})} 
:\! e^{i\alpha \theta(w,\bar{w})}\!:
+ \cdots
, \,\,
\\
&&
:\! e^{i\alpha\theta(z,\bar{z})}\!: \, 
  :\! e^{-i\alpha\theta(0,0)}\!: \,
\nonumber \\ && \hspace*{1cm}{}
= 
\frac{1}{|z|^{\alpha^2}}
  + \frac{\alpha}{|z|^{\alpha^2}} 
  \left(z J_\rho - \bar{z} \bar{J}_\rho \right)
+\frac{2i}{|z|^2} 
   \left(\partial_z \partial_{\bar{z}} \theta\right) 
\nonumber \\ && \hspace*{1.5cm} {}
+\frac{i\alpha}{2|z|^{\alpha^2}}
  \left[
     z^2  \left(\partial_z^2 \theta\right) 
     + \bar{z}^2  \left(\partial_{\bar{z}}^2 \theta\right) 
  \right]
\nonumber \\ && \hspace*{1.5cm} {}
+\frac{\alpha^2}{2|z|^{\alpha^2}}
  \left[
       z^2   : \! J_\rho ^2 \! :
     + \bar{z}^2  :\! \bar{J}_\rho^2 \! :
  \right]
\nonumber \\ && \hspace*{1.5cm} {}
-\frac{\alpha^2}{|z|^{\alpha^2-2}}
     J_\rho \, \bar{J}_\rho
+ \cdots ,
\\
&&
:\! e^{i\alpha\theta(z,\bar{z})}\!: \, 
  :\! e^{i\beta\theta(0,0)}\!: \,
= 
\frac{1}{|z|^{-\alpha\beta}} :\! e^{i(\alpha+\beta) \theta}\!:
 + \cdots ,
\end{eqnarray}
\label{eq:ope}%
\end{subequations}
   where we have introduced U(1) currents:
  $J_\rho(z)\equiv i \partial_z \theta(z,\bar{z})$,
  $\bar{J}_\rho(\bar{z}) \equiv
     - i \partial_{\bar{z}} \theta(z,\bar{z})$,
  $J_\sigma(z)\equiv i \partial_z \phi(z,\bar{z})$, and
  $\bar{J}_\sigma(\bar{z}) \equiv
     - i \partial_{\bar{z}} \phi(z,\bar{z})$.
The parameters $\alpha$ and $\beta$ ($\alpha + \beta \neq 0$) 
   in the vertex operator are 
   numerical constants which determine the scaling dimension.
In deriving the above OPE's, we have used the Wick theorem and
   the correlators: 
   $\langle \theta_+(\bar{z}) \, \theta_+(\bar{\omega})\rangle
      = -\frac{1}{2} \ln(\bar{z}-\bar{\omega})$,
   $\langle \theta_-(z) \, \theta_-(\omega)\rangle
      = -\frac{1}{2} \ln(z-\omega)$, and
   $\langle \theta(z,\bar{z}) \, \theta(\omega,\bar{\omega})\rangle
      = -\ln|z-\omega|$.
From Eq.\ (\ref{eq:ope}), one finds
\begin{subequations}
\begin{eqnarray}
\quad
&& \hspace*{-1.5cm}
\left[J_\rho(z) \, \bar{J}_\rho(\bar{z})\right] 
\,
\left[J_\rho(0) \, \bar{J}_\rho(0)\right] 
\nonumber \\
&=&
\frac{1}{|z|^4}
+ \frac{1}{|z|^4}
\left(z^2 : \!  J_\rho^2 \! : +  \bar{z}^2  :\! \bar{J}_\rho^2 \! :
  \right)
+\cdots, 
\label{eq:ope1}
\\
&& \hspace*{-1.5cm}
\left[J_\rho(z) \, \bar{J}_\rho(\bar{z})\right] 
 \, :\! \cos \alpha \theta(0,0) \!: 
\nonumber \\
&=&
  -\frac{\alpha^2}{4 |z|^2} \, :\! \cos \alpha \theta \! : 
+ \cdots,
\\
&& \hspace*{-1.5cm}
\left[J_\rho(z) \, \bar{J}_\rho(\bar{z})\right] 
 :\! \sin \alpha \theta(0,0) \!: 
\nonumber \\
&=&
  -\frac{\alpha^2}{4 |z|^2} \, :\! \sin \alpha \theta \! : 
+ \cdots,
\\
&& \hspace*{-1.5cm}
:\! \cos \alpha\theta(z,\bar{z})\!: \, :\!\cos \alpha\theta(0,0)\! :
\nonumber \\
&=&
\frac{1}{2|z|^{\alpha^2}}
+ \frac{\alpha^2}{|z|^{\alpha^2}}
\left(z^2 : \! J_\rho^2 \! : +  \bar{z}^2  :\! \bar{J}_\rho^2 \! :
  \right)
\nonumber \\ && {}
-\frac{\alpha^2}{2|z|^{\alpha^2-2}} \,
    J_\rho \, \bar{J}_\rho
\nonumber \\ && {}
+\frac{1}{2} |z|^{\alpha^2} :\!\cos 2\alpha\theta\! :
+ \cdots , \quad
\\
&& \hspace*{-1.5cm}
:\! \cos \alpha\theta(z,\bar{z})\!: \, :\!\cos \beta\theta(0,0)\! :
\nonumber \\
&=&
\frac{1}{2|z|^{\alpha\beta}} \, :\! \cos[(\alpha-\beta)\theta] \! :
\nonumber \\ && {}
+ \frac{1}{2 |z|^{-\alpha\beta}} :\! \cos[(\alpha+\beta)\theta] \! :
+\cdots .
\end{eqnarray}
\end{subequations}
Exchanging $\theta\to \phi$ and $\rho \to \sigma$ yields
  the OPE's for spin phase fields.

Expanding the action in powers of coupling constants and
   integrating out short-distance parts, we obtain the scaling
   equations,
\begin{eqnarray}
\frac{d}{dl} G_\Delta
 \!\!  &=& \!\! {}
G_\Delta \left(1+\frac{1}{2} \, G_\rho - \frac{1}{2} \, G_\sigma
    - G_c  - G_s
\right. \nonumber \\ && {} \left.
    - \frac{1}{2} \, G_{cs}
    - \frac{1}{4} \, G_{\rho s}
    - \frac{1}{4} \, G_{c\sigma}
    - \frac{1}{8} \, G_{\rho\sigma}
\right)
, \label{eq:RG-a}
\\
\frac{d}{dl} G_\rho 
 \!\!  &=& \!\! {}
   + \frac{1}{4} \, G_\Delta^2 + 2 \, G_c^2 + G_{cs}^2 +  G_s \, G_{\rho s}
,
\label{eq:RG-b}
\\
\frac{d}{dl} G_\sigma 
 \!\!  &=& \!\! {}
   - \frac{1}{4} \, G_\Delta^2  - 2 \, G_s^2 - G_{cs}^2 - G_c \, G_{c\sigma}
,
\\
\frac{d}{dl} G_c 
 \!\!  &=& \!\! {}
    -\frac{1}{4} \, G_\Delta^2
    + 2 \, G_\rho\, G_c
    - \left( G_s + G_{\rho s} \right) G_{cs}
,
\\
\frac{d}{dl} G_s 
 \!\!  &=& \!\! {}
   - \frac{1}{4} \, G_\Delta^2 - 2 \, G_\sigma \, G_s
   -  \left( G_c + G_{c\sigma} \right)  G_{cs}
,
\\
\frac{d}{dl} G_{cs} 
 \!\!  &=& \!\! {}
     -\frac{1}{4} \, G_\Delta^2 
    - 2 \left(1 - G_\rho 
         +  G_\sigma  + G_{\rho \sigma}\right)    G_{cs}
\nonumber \\ && {}
      - 2 \, (G_c + G_{c\sigma}) (G_s + G_{\rho s})
,
\\
\frac{d}{dl} G_{\rho s} 
 \!\!  &=& \!\! {}
    - \frac{1}{4} \, G_\Delta^2
    -2 \left(1 + G_\sigma \right) G_{\rho s}
    + 2 \, G_{\rho} \, G_s
\nonumber \\ && {}
    - 4 \, ( G_c +  G_{c\sigma} ) \, G_{cs}
    - 2 \, G_s \, G_{\rho \sigma}
,
\\
\frac{d}{dl} G_{c\sigma} 
 \!\!  &=& \!\! {}
    - \frac{1}{4} \, G_\Delta^2
    - 2 \left(1 - G_\rho \right) G_{c \sigma}
    - 2 \, G_{\sigma} \, G_c
\nonumber \\ && {}
    - 4 \, (G_s + G_{\rho s} ) \, G_{cs}
    - 2 \, G_c \, G_{\rho \sigma}
,
\\
\frac{d}{dl} G_{\rho \sigma} 
 \!\!  &=& \!\! {}
    - \frac{1}{4} \, G_\Delta^2 
    - 2 \, G_{\rho \sigma}
    + 2 \, G_\rho\, G_\sigma
    - 4 \, G_{cs}^2
\nonumber \\ && {}
    - 4 \, G_c \, G_{c\sigma} -4 \, G_s \, G_{\rho s} 
.
\label{eq:RG-e}
\end{eqnarray}
Here we note that the number of the RG equations
   can be reduced due to the spin-rotational SU(2) symmetry.
To show this point more transparently,
   we introduce $X(l)$, $Y(l)$, and $Z(l)$ by
   $X(l)=G_\sigma(l)-G_s(l)$,
   $Y(l)=G_{cs}(l)-G_{c\sigma}(l)$, and
   $Z(l)=G_{\rho s}(l)-G_{\rho \sigma}(l)$.
Their RG equations are obtained
   from Eqs.\ (\ref{eq:RG-b})--(\ref{eq:RG-e}) as
\begin{subequations}
\begin{eqnarray}
\frac{d}{dl} X
  &=& {}
    2 \, G_s \, X  + (G_c-G_{cs}) \, Y
,
\\
\frac{d}{dl} Y
  &=& {}
   2 \, ( - 1 + G_\rho + G_s + G_{\rho s}) \, Y
\nonumber \\ && {} 
   + 2 \, (G_c - G_{cs})  (X-Z)
, \\
\frac{d}{dl} Z
  &=& {}
     -2 \, (1 - G_s) \, Z
     -2 \, (G_\rho + G_{\rho s}) \, X
\nonumber \\ && {} 
     -4 \, (G_c - G_{cs}) \, Y
.
\end{eqnarray}
\end{subequations}
One immediately finds that,
  if $X(0)=Y(0)=Z(0)=0$, 
  they vanish for all $l$,
  i.e., $X(l)=Y(l)=Z(l)=0$.
This implies that 
   $G_\sigma(l)=G_s(l)$,
   $G_{cs}(l)=G_{c\sigma}(l)$, and
   $G_{\rho s}(l)=G_{\rho \sigma}(l)$, 
   which are nothing but the constraints on the coupling constants
   due to the spin-rotational SU(2) symmetry.
In this case, we can set
   $G_\sigma(l)=G_s(l)$, $G_{c\sigma}(l)=G_{cs}(l)$, and
   $G_{\rho \sigma}(l)=G_{\rho s}(l)$
   in the RG equations (\ref{eq:RG-a})--(\ref{eq:RG-e}).
Then the RG equations are given by 
   Eqs.\ (\ref{eq:RGf-b})--(\ref{eq:RGf-e}).
The RG equations for the 1D EHM without the staggered site potential
   are obtained by setting $G_\Delta(l)=0$,
   Eqs.\ (\ref{eq:Grho})--(\ref{eq:Grhos}).

The RG equations can also be obtained in the presence of the
   bond dimerization in a similar way.

\end{document}